\documentclass[sn-nature]{sn-jnl}

\usepackage{graphicx}%
\usepackage{multirow}%
\usepackage{amsmath,amsfonts}%
\usepackage{amsthm}%
\usepackage{mathrsfs}%
\usepackage[title]{appendix}%
\usepackage{xcolor}%
\usepackage{textcomp}%
\usepackage{manyfoot}%
\usepackage{booktabs}%
\usepackage{algorithm}%
\usepackage{algorithmicx}%
\usepackage{algpseudocode}%
\usepackage{listings}%
\usepackage{natbib}

\usepackage{glossaries}
\newacronym{rq1}{\textit{RQ 1}}{Research Question 1}
\newacronym{rq2}{\textit{RQ 2}}{Research Question 2}
\newacronym{rq3}{\textit{RQ 3}}{Research Question 3}
\newacronym{rq4}{\textit{RQ 4}}{Research Question 4}
\newacronym{eth}{\textit{ETH}}{Ether}
\newacronym{sc}{\textit{SC}}{Smart Contract}
\newacronym{evm}{\textit{EVM}}{Ethereum Virtual Machine}
\newacronym{dos}{\textit{DoS}}{Denial-of-Service}
\newacronym{dapp}{\textit{dApp}}{Decentralized Application}
\newacronym{api}{\textit{API}}{Application Programming Interface}
\newacronym{solc}{\textit{SOLC}}{Solidity Compiler}
\newacronym{smt}{\textit{SMT}}{Satisfiability Modulo Theories}
\newacronym{cfg}{\textit{CFG}}{Control Flow Graph}
\newacronym{gcg}{\textit{GCG}}{Gas Consumption Graph}
\newacronym{gastap}{\textit{GASTAP}}{Gas-aware Smart Contract Analysis Platform}
\newacronym{gasol}{\textit{GASOL}}{Gas Analysis and Optimization for Ethereum Smart Contracts}
\newacronym{tdge}{\textit{TDGE}}{Trace-based Dynamic Gas Estimator}
\newacronym{js-evm}{\textit{JS-EVM}}{JavaScript EVM}
\newacronym{rgum}{\textit{RGUM}}{Recent Gas Usage Model}
\newacronym{rgum-mean}{\textit{RGUM}$_{mean}$}{Recent Mean-Based Gas Usage Model}
\newacronym{rgum-median}{\textit{RGUM}$_{median}$}{Recent Median-Based Gas Usage Model}
\newacronym{rgum-max}{\textit{RGUM}$_{max}$}{Recent Maximum-Value-Based Gas Usage Model}
\newacronym{rgum-min}{\textit{RGUM}$_{min}$}{Recent Minimum-Value-Based Gas Usage Model}
\newacronym{acc}{\textit{ACC}}{General Accuracy}
\newacronym{acc+}{\textit{ACC}$_{+}$}{Positive Accuracy}
\newacronym{acc-}{\textit{ACC}$_{-}$}{Negative Accuracy}
\newacronym{ape}{\textit{APE}}{Absolute Percentage Error}
\newacronym{mape}{\textit{MAPE}}{Mean Absolute Percentage Error}
\newacronym{mape+}{\textit{MAPE}$_{+}$}{Positive Mean Absolute Percentage Error}
\newacronym{mape-}{\textit{MAPE}$_{-}$}{Negative Mean Absolute Percentage Error}
\newacronym{bgce}{\textit{BGCE}}{Baseline Gas Cost Estimator}
\newacronym{bgce-random}{\textit{BGCE}$_\textit{random}$}{Random Baseline Gas Cost Estimator}
\newacronym{bgce-default}{\textit{BGCE}$_\textit{default}$}{Default Baseline Gas Cost Estimator}
\newacronym{pe}{\textit{PE}}{Perfect Estimator}
\newacronym{ks}{\textit{KS}}{Kolmogorov-Smirnov}

\newcommand{\ttt}[1]{\texttt{#1}}
\newcommand{\ming}[1]{\textit{min}_{g}}

\usepackage{listings, xcolor}

\definecolor{verylightgray}{rgb}{.97,.97,.97}

\lstdefinelanguage{Solidity}{
	keywords=[1]{anonymous, assembly, assert, balance, break, call, callcode, case, catch, class, constant, continue, constructor, contract, debugger, default, delegatecall, delete, do, else, emit, event, experimental, export, external, false, finally, for, function, gas, if, implements, import, in, indexed, instanceof, interface, internal, is, length, library, log0, log1, log2, log3, log4, memory, modifier, new, payable, pragma, private, protected, public, pure, push, require, return, returns, revert, selfdestruct, send, solidity, storage, struct, suicide, super, switch, then, this, throw, true, try, typeof, using, msg.value, view, while, with, addmod, ecrecover, keccak256, mulmod, ripemd160, sha256, sha3}, 
	keywordstyle=[1]\color{blue}\bfseries,
	keywords=[2]{address, bool, byte, bytes, bytes1, bytes2, bytes3, bytes4, bytes5, bytes6, bytes7, bytes8, bytes9, bytes10, bytes11, bytes12, bytes13, bytes14, bytes15, bytes16, bytes17, bytes18, bytes19, bytes20, bytes21, bytes22, bytes23, bytes24, bytes25, bytes26, bytes27, bytes28, bytes29, bytes30, bytes31, bytes32, enum, int, int8, int16, int24, int32, int40, int48, int56, int64, int72, int80, int88, int96, int104, int112, int120, int128, int136, int144, int152, int160, int168, int176, int184, int192, int200, int208, int216, int224, int232, int240, int248, int256, mapping, string, uint, uint8, uint16, uint24, uint32, uint40, uint48, uint56, uint64, uint72, uint80, uint88, uint96, uint104, uint112, uint120, uint128, uint136, uint144, uint152, uint160, uint168, uint176, uint184, uint192, uint200, uint208, uint216, uint224, uint232, uint240, uint248, uint256, var, void, ether, finney, szabo, wei, days, hours, minutes, seconds, weeks, years},	
	keywordstyle=[2]\color{teal}\bfseries,
	keywords=[3]{block, blockhash, coinbase, difficulty, gaslimit, number, timestamp, msg, data, gas, sender, sig, msg.value, now, tx, gasprice, origin},	
	keywordstyle=[3]\color{violet}\bfseries,
	identifierstyle=\color{black},
	sensitive=false,
	comment=[l]{//},
	morecomment=[s]{/*}{*/},
	commentstyle=\color{gray}\ttfamily,
	stringstyle=\color{red}\ttfamily,
	morestring=[b]',
	morestring=[b]"
}

\begin{document}

\title[Demystification and Near-perfect Estimation of Minimum Gas Limit and Gas Used for Ethereum Smart Contracts]{Demystification and Near-perfect Estimation of Minimum Gas Limit and Gas Used for Ethereum Smart Contracts}


\author*[1]{\fnm{Danilo} \sur{Rafael de Lima Cabral}}\email{drlc@cin.ufpe.br}

\author[2]{\fnm{Pedro} \sur{Antonino}}\email{pedro@tbtl.com}

\author[1]{\fnm{Augusto} \sur{Cezar Alves Sampaio}}\email{acas@cin.ufpe.br}

\affil*[1]{\orgdiv{Centro de Informática}, \orgname{Universidade Federal de Pernambuco}, \orgaddress{\city{Recife}, \state{Pernambuco}, \country{Brazil}}}

\affil[2]{\orgname{The Blockhouse Technology Limited}, \orgaddress{\city{Oxford}, \country{UK}}}


\abstract{
The Ethereum blockchain has a \emph{gas system} that associates operations with a cost in gas units. 
Two central concepts of this system are the 
\emph{gas limit} assigned by the issuer of a transaction and the \emph{gas used} by a transaction. The former is a budget that must not be exhausted before the completion of the transaction execution; otherwise, the execution fails.
Therefore, it seems rather essential to determine the \emph{minimum gas limit} that ensures the execution of a transaction will not abort due to the lack of gas. Despite its practical relevance, this concept has not been properly addressed.
In the literature, gas used and minimum gas limit are conflated.
This paper proposes a precise notion of minimum gas limit and 
how it can differ from gas used
by a transaction;
this is also demonstrated with a quantitative study on real transactions of the Ethereum blockchain. Another significant contribution is the proposition of a fairly precise estimator for each of the two metrics. Again, the confusion between these concepts has led to the creation of estimators only for the gas used by a transaction. 
We demonstrate that the minimum gas limit for the state of the Ethereum blockchain (after the block) $t$ can serve as a near-perfect estimation for the execution of the transaction at block $t + \Delta$, where $\Delta \leq 11$; the same holds for estimating gas used. 
These precise estimators can be very valuable in helping the users predict the gas budget of transactions and developers in optimising their smart contracts; over and underestimating gas used and minimum gas limit can lead to a number of practical issues. Overall, this paper serves as an important reference
for blockchain developers and users as to how the gas system really works.
}


\maketitle

\section{Introduction}\label{section:introduction}

Ethereum is an open-source, decentralized, distributed computing platform \cite{Buterin:2014}, designed based on the blockchain technology originally introduced with Bitcoin \cite{Nakamoto:2008}. However, while Bitcoin aims to use the security of its immutable ledger to facilitate the exchange of coins among its users, Ethereum is proposed as a general-purpose global computer, in the sense that it enables developers to build and run \glspl{dapp} on its blockchain that maintains a single and globally accessible state that can be modified by running transactions related to two types of account: an externally owned account, also known as a user account, stores and transacts Ethereum's native currency, \gls{eth}; and a \gls{sc} account, a contract account, acts as a repository for code, typically written in a Turing-complete high-level programming language (such as Solidity \cite{Solidity}) and stored in \gls{evm}~\cite{Antonopoulos:2018} bytecode form. In a blockchain, the history of the system is captured by a chain of blocks, each of which contains a sequence of transactions. The single and global state of the blockchain is obtained by executing the transactions in the order induced by this history.

Although Turing completeness enables the development of more complex and versatile applications, it also introduces the challenge of dealing with infinite computations. To prevent coding errors or malicious attacks from causing \glspl{sc} to run their code indefinitely, Ethereum has incorporated a \emph{gas system}. Each transaction in the Ethereum blockchain has a gas limit element which is used by the issuer of the transaction to allocate an \emph{execution budget} that the transaction is allowed to use. Each computational operation on the Ethereum blockchain is associated with \emph{execution cost} measured in gas units. These units have a monetary value as they are worth some amount of \textit{Wei} ($10^{-18}$ \gls{eth}) \cite{Bouraga:2020}, and the issuer of a transaction pays for this budget before a transaction is executed. As operations for the transaction are executed, the initial budget is deducted. If the budget reaches zero before completion, the transaction execution is invalidated (i.e. its effects on the blockchain state are discarded); the issuer is not reimbursed despite the invalidation of the transaction as the effort related to the execution of the transaction has taken place and so it must be accounted for. These executions are said to trigger an \emph{out-of-gas exception} and, put simply, issuers for these transactions have paid for nothing. If there is a (positive) remaining budget, it is (converted back to some amount of \textit{Wei} and) given back to the issuer. The system also reports how many gas units have been consumed by a transaction in the blockchain; the receipt of a transaction contains such a gas used field.

The existence of a gas system poses an additional challenge for developers and users of the blockchain. Developers need to understand the consumption of gas by their contract to make them as gas-efficient as possible, and users must be able to set appropriate gas limits for transactions so they do not risk running out of gas. Therefore, developers would want to reliably estimate the gas used by their contracts and users the gas limit of the contracts they want to interact with. Such reliable estimations would offer many benefits. They could help users and developers understand the gas costs associated with contracts and transactions; the lack of predictability of such costs is one of the main obstacles to the widespread adoption of \glspl{dapp} \cite{AlShamsi:2022}. They could help to identify and optimise code patterns that consume more gas than necessary \cite{Marchesi:2020}. Finally, they could help to verify security risks associated with excessive gas consumption \cite{Grech:2018}. For instance, one security issue is \gls{dos} attacks which may exploit \glspl{sc} functions that do not adequately handle out-of-gas conditions \cite{Grech:2018}. 

The complexity of the gas system is even more prominent when the question of ``How to set the gas limit?'' arises. Arguably, the most appropriate value would be to set it to the smallest number of gas units such that the transaction executes without reverting due to an out-of-gas exception; we call this value \emph{the minimum gas limit}. If the gas limit is set to a value smaller than the minimum gas limit, the transaction will be reverted with an out-of-gas exception and the issuer will incur a financial loss. If the gas limit is set to a value higher than the minimum gas limit, the producer of the block that this transaction will be part of (who has the control over how transactions are ordered in the block) has an incentive to place the transaction is a way (i.e. in a position within the block such) that it consumes the extra gas units.

Intuitively, for a given transaction, one would expect the minimum gas limit and the gas used to coincide; after all, the budget necessary for a successful execution should correspond to the gas units used in the execution. However, in practice, that is not how the Ethereum \emph{gas system} works. It has peculiarities that, in many practical instances, drive the minimum gas limit away from the gas used for a transaction; typically, the minimum gas limit is greater than the gas used. This incorrect intuition has confused practitioners and academics alike. For instance, some academic papers wrongly (and implicitly) conflate these two concepts~\cite{Liu:2020,Zarir:2021}.

One of the main contributions of this paper is to propose an explicit definition for minimum gas limit and to 
emphasise
the difference between minimum gas limit and gas used. In Section \ref{section:minimum-gas-limit-vs-gas-used}, we present a very detailed exposition of the cases in which the calculations for the minimum gas limit and gas used diverge. This section demonstrates that these two concepts are not the same and should serve as a guide to the community to clarify the difference between them.

The other main contribution of this paper is to understand how these two metrics behave in practice and propose precise estimators for both of them. In Section~\ref{section:evaluation}, we analyse a period of the history of the Ethereum blockchain (Ethereum Mainnet's \textit{Bellatrix} fork) to understand their behaviour then. More precisely, we conduct an empirical study to address the following research questions. We use $t$ to stand for a block in the history of the chain and $\Delta$ an interval in number of blocks.

\textbf{\gls{rq1}}: \textbf{Under what circumstances is the minimum gas limit of a transaction at block $t$ expected to be a near-perfect estimation at block $t + \Delta$?}

\textbf{\gls{rq2}}: \textbf{Under what circumstances is the gas used by a transaction at time $t$ expected to be a near-perfect estimation at block $t + \Delta$?}

\textbf{\gls{rq3}}: \textbf{Are there significant differences between the gas used and minimum gas limit of Ethereum \gls{sc} call transactions?}

Intuitively, \gls{rq3} quantitatively examines the difference between the minimum gas limit and the gas used in practice; this quantitative assessment complements the qualitative analysis in Section~\ref{section:minimum-gas-limit-vs-gas-used}. Given an estimation of minimum gas limit at block $t$, \gls{rq1} examines how accurate is this estimation at block $t + \Delta$; \gls{rq2} follows the same pattern to examine gas used. 

In the context of \gls{rq3}, our studies demonstrate statistically significant differences between gas used and minimum gas limit for transactions. This result demonstrates that, in practice, transactions often fall into the cases which drive the values for minimum gas limit and gas used apart. Hence, developers and users should be aware of and expect such discrepancies when dealing with real-world smart contracts. As for \gls{rq1}, our analysis demonstrates that the calculation of the minimum gas limit of a transaction considering the state of the block $t$ serves as a near-perfect estimation (i.e., an estimate whose expected loss of precision is statistically insignificant) for the state of the block $t+\Delta$, when $\Delta \leq 11$. Similarly, in the context of \gls{rq2}, the calculation of gas used for the block $t$ serves as a near-perfect estimation for $t+\Delta$, for $\Delta \leq 11$. Even as $\Delta$ increases, the estimations remain reliable; of course, there is a degradation in the precision of the estimations as $\Delta$ increases. We also identify two classes of transactions that are defined based on how dynamic (i.e. their execution depends on the state of the blockchain) they are and analyse how these estimations vary between them. Thus, our study shows that calculations of minimum gas limit and gas used can be considered very reliable estimators within a reasonable time-bound. Additionally, these calculations can be performed by functions that are part of most Ethereum clients (programs to take part in the Ethereum blockchain). The function \textit{eth\_estimateGas()} calculates the minimum gas limit of a transaction for a given blockchain state whereas \textit{debug\_traceCall()} can be used to calculate the gas used for a specific transaction and blockchain state~\cite{Gethdoc, Paritydoc, Netherminddoc, Erigondoc,Ethereumdoc}.

We are unaware of any work that proposes or discusses the concept of minimum gas limit, let alone one that introduces the differences between this value and the gas used. We were able to find only a few works that systematically address the challenges related to gas estimation in Ethereum transactions. These works propose estimators for the gas used and suggest that this estimation be used to set the gas limit of the transaction, that is, these works implicitly conflate the minimum gas limit and the gas used. Liu et al. \cite{Liu:2020} categorise gas estimation methodologies into online and offline. According to this categorisation, offline estimators analyse the source code of \glspl{sc} to suggest the highest expected gas used for each of its functions \cite{Albert:2020}. On the other hand, online estimators predict the gas used of transactions based on data extracted from the Ethereum network or from a local blockchain operating under the same protocol \cite{Ma:2022}. A known issue with offline gas estimators is their inability to effectively handle functions that depend on state-based control flow, as they rely solely on static code analysis \cite{Li:2020}. Online estimators have to contend with the (usually complex) setup of a blockchain infrastructure from which they extract information. The estimators which we propose, which distinguish between gas used and minimum gas limit, are online and we rely on the fact that we use functions that are already available in Ethereum clients to alleviate the drawback of having to set up a blockchain infrastructure. Moreover, in the empirical study that we carry out, we compare our estimators with other online estimators (corrected to distinguish between gas used and minimum gas limit) and demonstrate that our strategy is statistically better than those.

Section~\ref{section:background} briefly introduces the main concepts of blockchains, \glspl{sc}, and the Ethereum gas system. Section~\ref{section:minimum-gas-limit-vs-gas-used} defines and explains the differences between the minimum gas limit and the gas used in an Ethereum transaction. To answer the research questions posed in this paper, in Section~\ref{section:evaluation} an experimental design and the related results are presented.  
Section \ref{section:related-work} discusses works related to the topic of our research. The final section
summarises our conclusions and outlines open areas for future research.

\section{Background} \label{section:background}

In this section, we 
succinctly introduce blockchains and \glspl{sc} (with a focus on Ethereum and Solidity) and the gas system used by Ethereum. 

\subsection{Blockchains} \label{subsection:blockchains}
A blockchain is a decentralised transaction-processing system. The blockchain participants issue transactions representing actions that they want to perform within the system --- such as transferring coins to other participants --- and a group of (blockchain maintainer) agents decides (i.e. reaches a consensus) on the validity of the transactions being processed and the order in which they are processed. These agents decide on a sequence of transactions that are aggregated into a block and they also decide on the sequence (i.e. chain; hence the name blockchain) of blocks that form the (immutable\footnote{Final or stable are also synonyms that appear in the blockchain literature.}) history of the system. The sequencing of transactions within a block and of blocks within the history of the blockchain defines the order in which transactions are processed and only valid transactions are added to blocks in this history. Typically, a blockchain relies on a lottery mechanism to choose an agent to create a candidate block to be the next one added to the history; this agent selects a sequence of transactions from a pool of transactions sent by participants. Then, a consensus protocol eventually decides if this candidate block is indeed added to the history of the system. If the protocol decides against adding this block, another candidate block is generated and decided upon. This generation-and-decision process is perpetuated until eventually a block is added to the history. Thus, the responsibility of these agents is to ever extend this chain of blocks.

The blockchain system also has a state $\sigma$ that is modified as transactions are processed. A blockchain can also be seen as a distributed database where each of these agents hold a copy of the state (i.e. database or digital ledger) and they agree on its value. This agreement is achieved via the same consensus protocol above. If all of these agents agree on the history of the blockchain and the process of a transaction affects the state of the chain in a deterministic way, it is not difficult to see that they all hold the same state. Typically, the state $\sigma$ captures the information required to implement the blockchain's transaction logic. For instance, the state $\sigma$ can have a mapping $\sigma_b$ of accounts (given here by an integer) to a balance (also an integer) so that $\sigma_b[10] = 7$ indicates that the account $10$ has a balance of $7$ coins. For this type of state, one can have transactions that would cause a transfer of coins between accounts. For example, let $\sigma_b[10] = 7$ and $\sigma_b[8] = 0$, a transaction $T = \textit{transfer}(10,8,5)$, which designates a transfer of $5$ coins from account $10$ to account $8$, when processed from state $\sigma$ would lead to state $\sigma'$ where $\sigma'_b[10] = 2$ and $\sigma'_b[8] = 5$; we use $\Upsilon(\sigma,T) = \sigma'$ to represent the processing of the transaction $T$ by the blockchain from state $\sigma$ leading to state $\sigma'$.\footnote{We borrow this notation from the Yellow Paper \cite{Wood:2014}.}

Blockchains were invented as a means to support a fully decentralised digital currency with Bitcoin \cite{Nakamoto:2008}. The original problem that it was proposed to tackle was \emph{double spending}, namely, how to prevent the same digital coin from being used twice; this effectively prevents digital money from being improperly created. Since then, this technology has been extended to tackle generic transaction logic that goes beyond double spending. With the advent of \glspl{sc}, blockchains that support this technology allow for a program to define how a transaction is to be processed, namely, how it affects the state of the blockchain.

\subsection{Smart Contracts} \label{subsection:smart-contracts}

\glspl{sc} have afforded blockchains a great level of flexibility. Instead of being constrained to implementing basic coin transfers and preventing double spending, blockchains with \glspl{sc} can rely on a program to dictate how transactions addressed to that contract are to be processed. Ethereum was the first blockchain network to implement \glspl{sc}, and it is still the most popular \gls{sc} platform to this date. In this subsection, we discuss the capabilities of \glspl{sc} by focusing on their implementation within the Ethereum blockchain. We focus on the details that are relevant for our exposition in this paper. For a full account of its behaviour, one should consult~\cite{Wood:2014}.

An account in Ethereum represents an aggregation of (digital) assets in a way that is similar in nature to a bank account. It is identified by \emph{an address}: a 160-bit non-negative integer, often presented in hexadecimal base. For instance, the address 0 is given by \lstinline|0x0|; this representation shortens the 40 zeros that should follow the hexadecimal prefix \verb|0x|. Ethereum has two types of account: externally owned account and \gls{sc}. The first type, also known as a user account, exists solely to store and transact Ethereum's native currency (\gls{eth}). This type of account is associated with a balance in the state of the blockchain. For instance, we use $\sigma_b[\ttt{0x0}] = 10$ to denote that the address \ttt{0x0} has balance of $10$ \textit{Wei} ($10^{-18}$ \gls{eth}; the indivisible unit of Ethereum's currency). A participant in the blockchain can generate a pair of cryptographic keys to control the associated externally owned account account. A transaction digitally signed by the private key denotes that the participant owning that key has consented to that transaction and, hence, the system should process it. An \gls{sc} account, often referred to as a contract account, acts as a repository for code, typically written in a Turing-complete high-level programming language (such as Solidity \cite{Solidity}) and stored in \gls{evm} bytecode form. This type of account also has a balance, representing the digital assets it owns, and the code dictates how the digital assets are managed, unlike externally owned account accounts where external participants issue transactions to move funds about. For this type of account, its state is represented by a balance $\sigma_b$ but it also has code $\sigma_c$ and storage $\sigma_s$; the latter stores the state (i.e. sometimes referred to as the \emph{storage}) of the \gls{sc}. We illustrate the behaviour of such an account by introducing Solidity, arguably the most popular high-level language used to write \glspl{sc} targeting Ethereum.

A contract in Solidity is similar in nature to that of a \emph{class} in object-oriented languages with its analogues of attributes and methods. We introduce the main constructs of Solidity using the \lstinline|Wallet| contract in Figure~\ref{figure:wallet-contract}. It implements a toy \textit{wallet} where participants and other contracts can deposit and withdraw their assets. The \emph{member variables} of a contract define the structure of the (persistent) state (i.e. storage) of the contract. A contract instance is an object persisted in the blockchain. This example contract has a single member variable \lstinline|balances|, a mapping from addresses to 256-bit unsigned integers, which keeps track of the balance of Wei for each address managed by the wallet; the integer \lstinline|balances[addr]| gives the current balance for address \lstinline|addr|.

\begin{figure}
\begin{lstlisting}[language=Solidity]
contract Wallet {
    mapping (address => uint) balances;

    function deposit() payable public {
        balances[msg.sender] = balances[msg.sender] + msg.value;
    }

    function withdraw(uint value) public {
        require(balances[msg.sender] >= value);
        bool ok = msg.sender.send(value);
        assert(ok);
        balances[msg.sender] = balances[msg.sender] - value;
    }
}
\end{lstlisting}
\caption{Wallet example contract.}
\label{figure:wallet-contract}
\end{figure}

\textit{Public functions} describe the operations offered by the contract, like methods in classes. The contract in Figure~\ref{figure:wallet-contract}
has two public functions: \lstinline|deposit| and \lstinline|withdraw|. The first one transfers Ether to the \lstinline|Wallet| contract, and the second withdraws Ether to the caller of the function.
In Solidity, functions have the implicit argument \lstinline|msg.sender| designating the caller's address, and \lstinline|payable| functions have the implicit \lstinline|msg.value| argument that depicts how much \emph{Wei} is being transferred, from caller to callee, with that function invocation; such a transfer is carried out implicitly by Ethereum. For instance,  when \lstinline|deposit| is called on an instance of \lstinline|ToyWallet|, the caller can decide on some amount \lstinline|amt| of Wei to be sent with the invocation. By the time the \lstinline|deposit| body is about to execute, Ethereum will already have carried out the transfer from the balance associated with the caller's address to that of the \lstinline{ToyWallet} instance, and \lstinline|amt| can be accessed via \lstinline|msg.value|. Note that, as mentioned, this balance is part of the blockchain's state rather than an explicit variable declared by the contract's code. One can programmatically access this implicit balance variable for address \lstinline|addr| with the command \lstinline|addr.balance|. 

Solidity's construct \lstinline{require(condition)}   aborts and reverts the execution of the function in question if \lstinline|condition| does not hold---even in the case of implicit Ether transfers. For instance, the \lstinline|require| statement in the function \lstinline|withdraw| requires the caller to have the funds they want to withdraw. The call \lstinline|addr.send(amount)| sends \lstinline|amount| Wei from the currently executing instance to address \lstinline|addr|; it returns \lstinline|true| if the transfer was successful, and \lstinline|false| otherwise.
The  \lstinline{assert} clause has a similar behaviour as that of \lstinline{require}; it has no effect if the condition holds and reverts the execution otherwise. However, whereas \lstinline{require} is used for input validation (for instance, parameter values), \lstinline{assert} is employed for checking internal errors and a postcondition violation. In the example, it is used to ensure that the  \lstinline|msg.sender.send(value)| statement must succeed, i.e. the \lstinline|value| must have been correctly withdrawn from \lstinline|Wallet| to \lstinline|msg.sender|. The final statement in this function updates the account balance of the caller (i.e. \lstinline|msg.sender|) in \lstinline|Wallet| to reflect the withdrawal.

\begin{figure}[t]
\centering
\begin{lstlisting}[language=C]
CALLVALUE 
PUSH1 0x0 
DUP1 
CALLER 
PUSH20 0xFFFFFFFFFFFFFFFFFFFFFFFFFFFFFFFFFFFFFFFF 
AND 
PUSH20 0xFFFFFFFFFFFFFFFFFFFFFFFFFFFFFFFFFFFFFFFF 
AND 
DUP2 
MSTORE 
PUSH1 0x20 
ADD 
SWAP1 
DUP2 
MSTORE 
PUSH1 0x20 
ADD 
PUSH1 0x0 
KECCAK256 
SLOAD 
ADD
\end{lstlisting}
\caption{EVM extract from deposit function.}
\label{figure:evm-deposit}
\end{figure}

In Ethereum, \glspl{sc} are stored and executed in \gls{evm}-bytecode format. So, a high-level language such as Solidity has to be compiled to \gls{evm} bytecode. This bytecode is executed by the \gls{evm}: a stack-based big-endian machine with a word size of 256 bits. An \gls{sc} is represented by \emph{a bytecode program}: a sequence of \gls{evm} instructions. An \gls{evm} instruction has an opcode and, optionally, some parameters, which can come from the code itself, the stack, etc. These opcodes have both an integer and a mnemonic representation. We present bytecode programs using the latter for the sake of readability. Most of the instructions are typical of bytecode languages but there are ones that are blockchain/Ethereum specific. We illustrate a fragment of a bytecode program in Figure~\ref{figure:evm-deposit}. The instruction \verb|CALLVALUE| places on the top of the execution stack the value for \lstinline|msg.value|, whereas \verb|CALLER| places \lstinline|msg.sender| on the top of the execution stack. The \verb|PUSH| instructions push a value onto the top of the stack; they range from \verb|PUSH1| to \verb|PUSH32|, each designed to push different lengths of data onto the stack. The instructions \verb|DUP1| and \verb|DUP2| duplicate the value of the first and second element in the stack and places it onto the top of the stack. \verb|AND| and \verb|ADD| have the usual logical and arithmetical meanings; they operate on the top two elements of the stack, consuming both and placing the resulting value on the top of the stack. The instruction \verb|MSTORE| stores a stack value in the execution memory whereas \verb|SLOAD| loads a storage value onto the stack. Finally, \verb|KECCAK256| calculates the Keccak-256 hash of a particular memory section; this hash calculation is used by Solidity to index member variables in storage. This program extract is part of the \gls{evm} bytecode generated for the \lstinline|deposit| function in Figure~\ref{figure:wallet-contract}. It computes the expression \lstinline|balances[msg.sender] + msg.value|. In this program, the instruction \verb|CALLVALUE| pushes \lstinline|msg.value| onto the top of the stack, and the sequence of instructions from the following \verb|PUSH1 0x0| to \verb|SLOAD| pushes the value of \lstinline|balances[msg.sender]| onto the top of the stack. This process involves calculating the storage position for this value, which is derived from the position of \lstinline|balances| and the value \lstinline|msg.sender|, using the Keccak-256 hash function, and loading this value from the storage to the stack. Then, these two values are added by the final \verb|ADD| instruction. A comprehensive presentation of the \gls{evm} instruction set can be found in \cite{Wood:2014}.

\subsection{Ethereum Transactions}
\label{subsection:ethereum-transactions}

Broadly speaking, there are three types of transactions in Ethereum: currency transfer, \gls{sc} creation, and \gls{sc} call. The first type simply moves Wei between externally owned account accounts. Transactions have elements \ttt{value}, \ttt{from}, \ttt{to} as in a currency transfer $T$ with $\ttt{value} = 10$ Wei, $\ttt{from} = \ttt{0x0}$ and $\ttt{to} = \ttt{0x1}$. We use this shorthand for addresses (as in $\ttt{0x0}$ and $\ttt{0x1}$) to represent addresses that are given, in practice, by a 40-digit long hexadecimal number. Also, let $\sigma$ be a blockchain state where $\sigma_b[\ttt{0x0}] = 20$ and $\sigma_b[\ttt{0x1}] = 5$. The processing of this transaction from that state leads to state $\Upsilon(\sigma,T) = \sigma'$  where $\sigma'_b[\ttt{0x0}] = 10$ and $\sigma'_b[\ttt{0x1}] = 15$. 

An \gls{sc} creation transaction $T$ with bytecode program $bc$ creates a new contract account with this associated code\footnote{The \gls{sc} creation behaviour is more intricate than that as it has an \textit{init} code element that is run once to initialise the storage of the \gls{sc}, in addition to the $bc$ component. This behaviour is not relevant, however, for our presentation here.}. Assuming that the contract was created at address \ttt{0x5}, the processing of this transaction would result in a state $\Upsilon(\sigma,T) = \sigma'$ where $\sigma'_c[\ttt{0x5}] = bc$; this code could be, for example, the compiled code for the contract \verb|Wallet| in Figure~\ref{figure:wallet-contract}. 

For an \gls{sc} call transaction, the element \ttt{to} designates the address of the contract being called. The transaction also contains a byte string (as the \ttt{data} element) to be passed as a parameter to the bytecode program. For Solidity programs, for example, this byte string designates the function that is being called and contains the parameters for that function. An \gls{sc} call transaction triggers the execution of code that typically updates the storage of that contract. For instance, let $T$ be an \gls{sc} call transaction that invokes \lstinline{deposit} on an instance of \verb|Wallet| at address $\ttt{to} = \texttt{0x5}$ with $\ttt{from} = \ttt{0x1}$ and $\ttt{value} = \ttt{10}$, and let $\sigma$ be an Ethereum state such that $\sigma_s[\ttt{balance}][\ttt{0x1}] = 0$, $\sigma_b[\ttt{0x1}] = 20$, and $\sigma_b[\ttt{0x5}] = 0$. Processing this transaction would lead to a state $\Upsilon(\sigma,T) = \sigma'$ such that $\sigma'_s[\ttt{balance}][\ttt{0x1}] = 10$, $\sigma'_b[\ttt{0x1}] = 10$, and $\sigma'_b[\ttt{0x5}] = 10$. Note that the transaction element \ttt{value} is accessed via \lstinline{msg.value} whereas the element \ttt{from} is accessed via \lstinline{msg.sender}. A(n) (external) transaction in Ethereum can only be initiated by an externally owned account, that is, for all these transactions the \ttt{from} element must be such an account. An \gls{sc} creation or call can also contain a \ttt{value} element that triggers a transfer of Wei to the target \gls{sc}, as shown in the example above. During the execution of an \gls{sc}, it may call other \glspl{sc} or even transfer currency to externally owned accounts---the \gls{evm} instruction set has opcodes implementing both of these behaviours.

Although Turing completeness enables the development of more complex and versatile applications, it also introduces the challenge of dealing with potentially unbounded loops. To prevent coding errors or malicious attacks from causing \glspl{sc} to run their code indefinitely, Ethereum has incorporated a gas system that associates the cost of each computational operation (i.e. \gls{evm} instructions) on the blockchain with a specific amount of \emph{gas units} (which costs \textit{Wei}) \cite{Bouraga:2020}.

\section{Unpacking \emph{Gas}: Minimum Gas Limit $\neq$ Gas Used} \label{section:minimum-gas-limit-vs-gas-used}

In this section, we define two gas concepts that are often confused and conflated---minimum gas limit and gas used---and explain their difference according to the Ethereum gas system \cite{Wood:2014}. The former corresponds to the minimum amount of gas necessary to execute a function without reverting, the latter corresponds to the actual gas consumed by the transaction. Although these are relatively simple and well-defined concepts, they are generally misused in the literature with consequent impact on estimation approaches. We also contribute with a discussion on estimating these values: what issues can arise if estimations are wrong and how one can estimate them using functions available in execution layer clients.

\subsection{Ethereum Gas System}
\label{section:ethereum-gas-system}

In Ethereum, a transaction is given a budget that is consumed as it is processed. We discuss only the consumption by a transaction executing code since simple transfers have a fixed amount of gas associated with them. Moreover, we do not investigate in this paper \gls{sc} creation transactions. We detail the most relevant elements of this process in this subsection; for a full account see the Yellow Paper~\cite{Wood:2014}.

A transaction has a gas limit $T_g$---a budget in units of gas---which is set by the sender of the transaction. The sender must pay for this budget up front, i.e., their balance must be sufficient to pay for this budget. Otherwise, the transaction is not deemed valid and will not be processed by the blockchain. We do not discuss how gas units are priced as these calculations are immaterial for this paper. What remains of the budget at the end of a transaction execution is returned to the sender.

The cost of a transaction $T$ is given by $g_\textit{cost}$, as follows: 
\begin{equation}
    g_{cost} = g_0 + g_\textit{exec}
\end{equation} 
where $g_0$ denotes $T$'s \emph{intrinsic cost}, and $g_\textit{exec}$ denotes $T$'s \emph{execution cost}.\footnote{For transactions that create \glspl{sc}, there is an additional component $g_c$ that gives the $T$'s \emph{code allocation cost} which is proportional to the size of the contract being created.}
The intrinsic cost is deducted before the transaction is properly executed. Thus, the gas limit $T_g$ must cover the intrinsic gas $g_0$, i.e. $g_0 \leq T_g$. Otherwise, the transaction is not considered valid and will not be processed by the blockchain. 

The intrinsic cost of $T$ is calculated as follows:
\begin{equation}
    g_0 = g_\textit{data} + g_\textit{create} + g_\textit{tx} + g_\textit{access}
\end{equation}
where $g_\textit{data}$ gives the cost of allocating data/code for the execution of the transaction;  $g_\textit{create}$ is either $32000$ gas units, if the transaction involves the creation of a contract, or $0$, otherwise; $g_{tx}$ is a (flat) constant transaction cost of $21000$ gas units; and $g_\textit{access}$ accounts for the allocation of the transaction data related to the account
and storage accesses. It is proportional to the size of the \textit{accessList} element\footnote{The accessList aims to optimize gas costs by specifying a list of addresses and storage keys that a transaction intends to access. This helps the \gls{evm} pre-load these addresses and storage slots, which can result in gas savings.} of a transaction.

Once the intrinsic cost is deducted, the remaining (execution) gas budget $g' = T_g - g_0$ funds the (code) execution of the transaction. The execution cost cannot exceed this budget so we have that $g_\textit{exec} \leq T_g - g_0$, and hence that: 
\begin{equation} \label{equation:cost-limit}
g_\textit{cost} \leq T_g
\end{equation}
The execution cost is derived from the \gls{evm} instructions performed in the execution of $T$. Each opcode has an associated cost of gas units reflecting its computational demands. For example, an addition operation (opcode \verb|ADD|) costs 3 gas units, whereas a jump operation (opcode \verb|JUMP|) costs 8 gas units; a full account of instructions and their cost can be found in the appendix of~\cite{Wood:2014}. In practice, the \gls{evm} does not calculate this cost \emph{per se}; instead, it updates (i.e. deducts) each instruction cost from $g'$ as the transaction is being executed. For instance, Figure~\ref{figure:execution-cost} presents an example of this behaviour where the initial execution budget is $T_g - g_0 = g' = 10,000$. The execution cost can be calculated by $t_\textit{exec} = T_g - g_0 - \hat{g}$ where $\hat{g}$ is the remaining execution budget $g'$ after execution has finished. For example, for the execution in Figure~\ref{figure:execution-cost}, we have that $t_\textit{exec} = 2109$. If during this process the budget reaches zero (an out-of-gas exception), the processing is aborted and the changes to the state of the blockchain that have been carried out during the execution of the transaction are reverted. Other types of exceptions may occur in the processing of a transaction leading to the same pattern of state-change reversal as gas-budget consumption.

\begin{figure}
\begin{lstlisting}[language=C]
    // g' = 10,000 
    PUSH1 0x1 // cost 3
    // g' = 9,997
    PUSH1 0x0 // cost 3
    // g' = 9,994
    SLOAD // cost 2100 (assuming cold storage load)
    // g' = 7,894
    ADD // cost 3
    // g' = 7,891
\end{lstlisting}
\caption{Example execution cost calculation.}
\label{figure:execution-cost}
\end{figure}

The sender of a transaction may be rewarded for clearing storage values in the blockchain. The \lstinline{SSTORE} instruction 
pops the value and the storage slot from the execution stack and writes the value to the specified slot. When this instruction is used to clear storage values, the refund value $A_r$ is updated; the precise calculation of this value is irrelevant to this paper but it can be found in \cite{Wood:2014}. At the end of the execution of the transaction, the refund value $R$ below is used to amortise the cost of the transaction. 
\begin{equation}\label{equation:refund}
    R = \textit{min}\left\{ \left\lfloor \frac{T_g - g_\textit{cost}}{5} \right\rfloor ,A_r \right\}
\end{equation}
The value of the refund is limited to 1/5 of the difference between the transaction budget ($T_g$) and the transaction cost ($g_{cost}$). In the particular case when $g_\textit{cost} = T_g$, we have that $R=0$.
Finally, the gas used by the transaction is calculated as follows.
\begin{equation}\label{equation:gas-used}
    g_\textit{used} = g_\textit{cost} - R 
\end{equation}

The sender of the transaction is reimbursed for the gas left: $T_g - g_\textit{used}$. For a transaction $T$ and a blockchain state $\sigma$, we use $\Upsilon^g(\sigma,T)$ to denote the associated $g_\textit{used}$. Note that the state $\sigma$ impacts this value as the instructions executed (and hence $g_\textit{exec}$) can be affected by it. Moreover, note that from Equations~\ref{equation:cost-limit} and~\ref{equation:gas-used}, we can deduce that: 
\begin{equation}
 g_\textit{used} \leq T_g   
\end{equation}

\subsection{Minimum Gas Limit versus Gas Used} \label{subsection:minimum-gas-limit-versus-gas-used}

In Ethereum, a transaction execution has a status code $z$ that indicates if the transaction effects have been committed ($z=1$) or reverted ($z=0$). Some opcodes trigger a revert, like \verb|REVERT| and \verb|INVALID|. Moreover, there are conditions such as running out of gas during the execution of an \gls{sc} or not having enough elements on the stack for the execution of a given instruction. All of these conditions lead the \gls{evm} to an exceptional halting state where the execution is reverted\footnote{The function $Z$ in \cite{Wood:2014} formalises these conditions.}. For a transaction $T$ and a state $\sigma$, we use the function $\Upsilon^z(\sigma,T)$ to denote the status code $z$ for the execution of this transaction from this state. A formal definition for the status code of a transaction execution is given in \cite{Wood:2014}.

The minimum gas limit is the smallest gas budget that can be given for a transaction leading to a committed execution. Given a transaction $T$ and a blockchain state $\sigma$, the minimum gas limit $\ming{}(\sigma, T)$ is formally calculated as follows. We use $T(g^*)$ to denote that $T$ has had its gas budget $T_g$ replaced by $g^*$. 

The function \textit{min} yields the minimum value in a set of integers; we assume that this set is non-empty for the sake of simplicity.
\begin{equation}
    \ming{}(\sigma, T) = \textit{min}~\{g \mid \Upsilon^z(\sigma,T(g)) = 1\}
\end{equation}

Note that this definition does not depend on the original gas limit assigned to the transaction ($T_g$) as $\ming{}$, in general, yields a transaction $T(\ming{}(\sigma, T))$ that is different from $T$, unless $T_g = \ming{}(\sigma, T)$. Hence, we cannot simply use the Equation~\ref{equation:gas-used} to derive (ii) $\Upsilon^g(\sigma,T) \leq \ming{}(\sigma, T)$; this equation compares $g_\textit{used}$ and $T_g$ for the same transaction whereas (ii) may involve two different transactions (which differ on their gas budget). We give a counterexample to illustrate this discrepancy.

\begin{figure}[h]
\begin{lstlisting}
function discrepancy() public {
    if (gasleft() < 30000) {
        // omitted operations that consumes 1000 gas
    } else {
        // ommitted operations that consumes 2000 gas
    }
}
\end{lstlisting}
\caption{A counterexample to $\Upsilon^g(\sigma,T) \leq \ming{}(\sigma, T)$.}
\label{figure:discrepancy}
\end{figure}

Let us assume that $T$ is a transaction that calls the \lstinline|discrepancy| function presented in Figure~\ref{figure:discrepancy}. Note that it is possible to setup the gas used by $T$ (with $T_g$ $\geq 30000$) such that it
includes the $2000$ gas cost of the \lstinline|else| branch\footnote{We assume that the omitted execution is so that a gas budget of $g_\textit{cost}$ makes the transaction execute successfully.}. On the other hand, considering only the $1000$ gas cost of the \lstinline|then| branch, the transaction $T(g_\textit{min})$ is setup with $1000$ gas limit.
 Hence, in this case, we have that $\Upsilon^g(\sigma,T) > \ming{}(\sigma, T)$. This illustrates, yet again, the dissimilarity between gas used and minimum gas limit.

We can, however, derive from Equation~\ref{equation:gas-used} that $\Upsilon^g(\sigma,T(\ming{}(\sigma, T))) \leq \ming{}(\sigma, T)$ as both sides now relate the same transaction; note that this equation now refers to the gas used by the transaction $T(\ming{}(\sigma, T))$ instead of $T$. We show in the following that, for some common cases, it is actually the case that: 
\begin{equation}
\Upsilon^g(\sigma,T(\ming{}(\sigma, T))) < \ming{}(\sigma, T)    
\end{equation} 
This expresses that the minimum gas limit is \emph{strictly} higher than the gas used. For the sake of simplicity, henceforth, we consider the case where $T_g = g_\textit{min}$ and so $\Upsilon^g(\sigma,T) = \Upsilon^g(\sigma,T(\ming{}(\sigma, T)))$.

The refund mechanism implemented by the \verb|SSTORE| instruction drives $\ming{}(\sigma, T)$ and $\Upsilon^g(\sigma,T)$ apart. As we have discussed in the previous section, the execution cost $g_\textit{cost}$ for a transaction must be met by the gas budget $T_g$ before it can be amortised by a potential refund $R$. Thus, the minimum gas limit required should be closer in nature to $g_\textit{cost}$ as opposed to $g_\textit{used}$, namely, $g_\textit{cost} = g_\textit{used} + R$ should be a better approximation for $\ming{}(\sigma, T)$. There are, of course, executions in which $R=0$ and so $g_\textit{cost} = g_\textit{used}$. Figure \ref{figure:refund} illustrates this divide between these two values. For a \lstinline{uint} member variable \verb|a| of the contract, if this function is called from a state where $\ttt{a} \neq 0$, it will contribute for a refund of $4800$ gas units; to be more precise, the \verb|SSTORE| implementing this storage write generates this refund. For this case, $\ming{}(\sigma, T) = gas_\textit{cost} = \Upsilon^g(\sigma,T) + 4800$. Writing to storage and having such a refund is very common in Ethereum.

\begin{figure}[h]
\begin{lstlisting}
function clear_a() public {
    a = 0;
}
\end{lstlisting}
\caption{Refund example: clearing member variable \lstinline{a}.}
\label{figure:refund}
\end{figure}

There are cases in which not even $g_\textit{cost}$ is a precise approximation for the minimum gas limit. For example, the \verb|SSTORE| instruction needs the gas budget to exceed a \emph{gas stipend} by 2300 units. This value is \emph{not consumed} by the instruction but an exceptional halting state is reached if the gas budget is at most this limit. Therefore, if an execution reaches such an instruction and the cost to finish this execution is $h \leq 2300$, the minimum gas limit would need to be higher than the execution cost to ensure that the gas budget available at this point is greater than $2300$ and then avoid an exception. Thus, the minimum gas limit would have to be increased by $2301 - h$ and so $\ming{}(\sigma, T) = g_\textit{cost} + 2301 - h$, assuming for simplicity that there is no refund and that this is the only discrepancy between cost and minimum gas. We illustrate this difference with the example in Figure~\ref{figure:warm-load}; again, we assume a \lstinline{uint} member variable \verb|a| of the contract. The \gls{evm} code generated for this function does not give any refund and the execution of this function from (and including) the \verb|SSTORE| instruction implementing the  \lstinline{a = x} assignment (until the end of its execution) costs $h=115$ gas units. Hence, for this specific example, $\ming{}(\sigma, T) = g_\textit{cost} + 2301 - 115 = g_\textit{cost} + 2286$. This may seem an uncommon example, but functions that end on a member variable assignment are likely to fall into this category, and such a code pattern is commonplace in Solidity contracts. 

\begin{figure}[h]
\begin{lstlisting}
function warmload_a() public {
    uint x = 4;
    uint y = 3;
    a = x;
    a = y;
}
\end{lstlisting}
\caption{Warm load example: second write leads to a warm load of \lstinline{a}. }
\label{figure:warm-load}
\end{figure}

In general, if an execution has to meet some budgetary constraint that is not reflected in the gas cost of the transaction, the delta between the budget required and the execution cost will affect how close the minimum gas limit and the gas cost are. More precisely, we call a budgetary constraint a condition on the remaining gas budget $g'$ at some point in the program, i.e. $g' > c$ for some gas value $c$, that if not met it will cause the execution to revert. If a non-reverting execution reaches a point with such a condition $g' > c$ and the cost for completing that execution is $h \leq c$, then it must be that
\begin{equation}
c - h + 1 \tag{i}
\end{equation}
separates the minimum gas limit and the gas cost; $\ming{}(\sigma, T) = g_\textit{cost} + c - h + 1$. In this statement, we assume that there is only one such budgetary constraint point in this execution for the sake of simplicity. In general, there could be more and these may have a compound effect on one another.
The gas stipend for \verb|SSTORE| is an example of this behaviour where $c = 2300$. In fact, the \gls{evm} instruction set allows one to programmatically create this type of constraint (choosing an arbitrary $c$) and behaviour. The \gls{evm} instruction \verb|GAS| places on top of the stack the remaining gas budget for the execution. Hence, one can use this value to control the flow of a bytecode program. We exemplify this sort of behaviour with the function in Figure~\ref{figure:gas-control-flow}. In this function, the remaining gas is returned by the built-in Solidity function \lstinline{gasleft()} (this function is implemented using the instruction \verb|GAS|) and the execution reverts if this budget is smaller than $c = 30000$ gas units. Considering the bytecode generated for this function, the cost from (and including) the \verb|GAS| instruction until the end of a non-reverting execution (i.e. skiping the revert if branch) is $h=31$ gas units. This means that, for this execution, applying (i) the difference between gas cost and minimum gas limit is $30000 - 31 + 1 = 29970$. There are real contracts deployed in the Ethereum network that implement this type of control-flow logic based on the remaining gas available.

\begin{figure}[h]
\begin{lstlisting}
function gasleft_control_flow() public {
    if (gasleft() < 30000) {
        revert();
    }
}
\end{lstlisting}
\caption{Gas-budget reverting dependency example.}
\label{figure:gas-control-flow}
\end{figure}

These examples illustrate how differences between the gas used, the gas cost, and the minimum gas limit arise in practice, through code fragments of real contracts. 

\subsection{Estimation of Minimum Gas Limit} \label{subsection:estimation-of-minimum-gas-limit}

The existence of a gas system poses an additional challenge for developers who want to create Ethereum \glspl{sc}. This is because, in addition to ensuring that their code works correctly and is computationally viable and secure, developers need to understand how this gas system works to make their \glspl{sc} as cost-efficient as possible. Associated with this need, it is important that developers are able to reliably estimate the minimum gas limit of their contracts to offer more transparency to users of applications that use such \glspl{sc} \cite{Wessling:2018}, identify and optimise code patterns that require more gas than necessary \cite{Marchesi:2020}, and verify security risks associated with excessive gas requirement \cite{Grech:2018}.

Estimating the minimum gas limit (or gas used) for a transaction is a difficult task given that the cost of an execution is associated with the \emph{dynamic} state of the blockchain. For instance, instructions like \verb|SSTORE| and \verb|SLOAD| have their gas cost calculated based on the values in the storage of the contract being executed and, moreover, their cost also depends on the execution itself; for instance, the gas cost of \verb|SSTORE| depends on whether it is writing to the storage address for the first time or not in this execution. So, one can only reliably calculate that, in general, if they know what is the state from which that transaction is executed. This is usually not the case when a participant sends a transaction to the blockchain network. 

Formally, let $g_\textit{min}$ be the minimum gas limit for transaction $T$ executing on a blockchain with state $\sigma$. The underestimation and overestimation of the minimum gas limit for $T$ and $\sigma$ occur, respectively, when $T_{g} < g_{\textit{min}}$ and $T_{g} > g_{\textit{min}}$. Both cases can lead to a range of issues.

Underestimation can lead to a wasteful consumption of gas: processing (leading to the use of gas) occurs but its effects are not committed --- in such a case, an execution is a costly no-op-like operation. The $T_{g}$ underestimation can arise in two ways:

\begin{enumerate}

    \item $T_{g} < g_{0}$: if the intrinsic gas cost of a transaction is not met by its gas budget, the transaction is not considered valid and it is not executed. Thus, there is no loss as far as the blockchain financial mechanisms. Nevertheless, the participant has to re-send the transaction, and so the resources used to generated the invalid transaction (i.e. energy/computing power) may be seen as a(n) (external) financial penalty incurred by the participant.
    \item $g_{0} \leq T_{g} < g_{\textit{min}}$: if the transaction's gas budget is smaller than the minimum gas limit but bigger than the intrinsic gas cost, the transaction is executed but its effects (if they exist) are reverted (by the definition of minimum gas limit), yet some amount of gas is consumed. There are two possibilities for this consumption:
    \begin{enumerate}
        \item if the execution has reached a \verb|REVERT| instruction, the amount of gas used until then is consumed;
        \item if the execution reaches an exceptional halting state (e.g. if it has no gas left to execute, or has reached a \verb|SSTORE| without at least 2301 left in the gas budget\footnote{Function $Z$ in \cite{Wood:2014} formalises such executions.}), all of the gas budget made available to the transaction, i.e. $T_g$, is consumed.
    \end{enumerate}
    Therefore, the punishment for these sorts of transactions are given within the financial system of Ethereum itself. Additionally, the same re-sending punishment is inflicted in the participant, if they want the effects of the transaction to be committed by a new transaction.
\end{enumerate}

Overestimating $T_g$ tends to be less harmful, in most cases the amount of unused gas is just returned to the sender of the transaction. However, perverse incentives are created by overestimation. The cases for overestimation are:

\begin{enumerate}
    \item $\textit{block}_g < T_g$: all blocks on the Ethereum network have a maximum limit of gas units that can be used by the transactions contained within them \cite{Wood:2014}. Although unlikely, a user may set a $T_g$ value that exceeds this limit, thereby making it impossible to include the transaction in the block.

    \item $g_\textit{from} < T_g$: this issue has a similar effect to what happens when $T_g < g_0$. This occurs because the \gls{evm} will only start executing $T$ if the sender has sufficient balance to cover the transaction expenses, 
    as specified in the previous subsection.

    \item $g_\textit{min} < T_g < g_\textit{top}$ (where $g_\textit{top} = \textit{min}~\{g_\textit{block}, g_\textit{from}\}$): if the budget is higher than the minimum gas limit, the remaining unused gas is returned to the sender. However, there is a subtle issue with such an overestimation: the sender is creating an incentive for the producer of the block containing this transaction to ``play" the state of the blockchain. Let us assume that the participant is expecting an execution from a state $\sigma_{exp}$ such that $\textit{min}_g(\sigma_{exp},T) = g_\textit{min\_exp}$ but they decide to give a budget $T_g > g_\textit{min\_exp}$. In this case, the participant is effectively creating an incentive $ict = T_g - g_\textit{min\_exp}$ for the producer of the block containing this transaction to manipulate the state $\sigma^*$ from which the transaction is processed to make the transaction use the extra $ict$ gas units. Block producers are rewarded for their work based on the amount of gas used by the transactions in the block being produced. Hence, they have an incentive to maximise that.
\end{enumerate}

Blindly trying to overestimate the gas budget of a transaction (by relying on the fact that the unused part of the budget is returned) can be a bad strategy. If the contract code contains a bug causing it to traverse its code through endless paths \cite{Marchesi:2020}, this type of contract flaw could result in significant financial losses.

\subsection{Calculating Gas Used and Minimum Gas Limit}
\label{subsection:calculating-and-estimating-gas-used-and-minimum-gas-limit}

Execution layer clients offer functions to measure the gas used by a transaction (\textit{debug\_traceCall}) and to calculate the minimum gas limit for a transaction (\textit{eth\_estimateGas}). In this subsection, we discuss how these functions operate, how they can be used to estimate the gas used and the minimum gas limit for a transaction, and some of their limitations. We describe here these functions as they have been implemented by Geth, a Go implementation of Ethereum~\cite{Geth}; there may be some discrepancies between client implementations. We describe only the elements of these functions that are relevant for our exposition. For instance, we omit parameters that are not relevant for the analysis that we conduct. For a full account on how these functions operate the reader is referred to the documentation of this client~\cite{Gethdoc}.  

The function \textit{debug\_traceCall}($B, T$) takes as parameters a block identifier $B$ and a transaction $T$ and execute this transaction in  a context where it is added on top (i.e. as the new last transaction) of block $B$. This function collects a detailed trace of this execution that includes the gas used, status code, etc. In this paper, we precisely define the projected behaviour we use of such a function. This is given by the function \textproc{TraceCall} that receives a blockchain state $\sigma$ (as opposed to a block identifier) and a transaction $T$ as parameters, and execute the transaction from that state. We are interested only in the status code of $(\Upsilon^z(\sigma,T))$ and the gas used by  ($\Upsilon^g(\sigma,T))$ the execution. So, these are the only two values that this function returns. Its precise definition is given by:
\begin{equation*}
    \textproc{TraceCall}(\sigma, T) = (\Upsilon^z(\sigma,T),\Upsilon^g(\sigma,T))
\end{equation*}
Note that if we have $\sigma_B$ as the state of the blockchain right after processing the last transaction of block $B$, \textproc{TraceCall}($\sigma_B, T$) carries out the same execution as \textit{debug\_traceCall}($B, T$), when projected to the output pair: status code and gas used.

The function \textit{eth\_estimateGas}($B, T$) also takes as parameters a block identifier $B$ and a transaction $T$ and tries to find the minimum gas limit for this transaction, again,  in the context where it is the last transaction of block $B$. This function performs a binary search trying to identify the smaller gas budget that enables the transaction to execute without reverting. We use the definition \textproc{EstimateGas} given in Algorithm~\ref{algorithm:estimate} to capture the precise behaviour of this function; once more, we use a blockchain state instead of a block identifier. We use $g_0$ to represent the intrinsic gas value for a transaction, $g_\textit{top} = \textit{min}~\{g_\textit{block}, g_\textit{from}\}$ represents the upper limit of our search range ($g_\textit{block}$ represents the gas budget of the block---a transaction in this block cannot exceed this value---and $g_\textit{from}$ represents the maximum amount of gas the sender of the transaction---address \lstinline{from}---can afford\footnote{This correspond to the maximum amount of gas units that can be purchased with the sender's balance, i.e. $\sigma_b[\ttt{from}]$. For our exposition, it does not matter how this value is calculated. It suffices to know that this is the maximum gas the sender can afford.}), 
$(x,y) := (v, w)$ represents a simultaneous assignment where $v$ is assigned to $x$ and $w$ to $y$ (the placeholder $\_$ can be used on the left-hand side of a simultaneous assignment to denote that the associated value is discarded), $\bot$ represents an \emph{uninitialised} value, and $T(g)$ is the transaction resulting after replacing $T_g$ with $g$ in transaction $T$. In Algorithm~\ref{algorithm:estimate}, the minimum gas limit $g_\textit{min}$ is updated as new values of $g \leq g_\textit{min}$ leading to a non-reverting execution (i.e. where $z = 1$) are found; the function \textproc{TraceCall} is used to carry this check. Variables $L$ and $H$ determine the interval where the binary search is looking for potential values for the minimum gas limit. 

\begin{algorithm}[h]
\begin{algorithmic}
\Function{EstimateGas}{$\sigma, T$}
    \State $L := g_0$
    \State $H, g := g_\textit{top}, g_\textit{top}$
    \State $\hat{g}_\textit{min} := \bot$
    \While{$L \leq H$}
        \State $(z,\_) :=$ \Call{TraceCall}{$\sigma,T(g)$}
        \If{$z = 1$}
            \State $\hat{g}_\textit{min} := g$
            \State $H := g - 1$
        \Else
            \State $L := g + 1$
        \EndIf
        \State $g := \lfloor(L + H) / 2\rfloor$
    \EndWhile
    \State \Return $\hat{g}_\textit{min}$
\EndFunction
\end{algorithmic}
\caption{Pseudocode for the estimation of the minimum gas limit for a given blockchain state $\sigma$ and transaction $T$.}
\label{algorithm:estimate}
\end{algorithm}

This binary-search approach does not always find the minimum gas limit. Let $[l_1,u_1],\ldots,[l_n,u_n]$ be the $n$ \emph{non-reverting gas (sub-)intervals} within $[g_0,g_\textit{top}]$ defining the gas budgets $g$ for which $\Upsilon^z(\sigma,T(g)) = 1$, namely, the transaction $T(g)$ leads to a non-reverting execution from state $\sigma$ if and only if $g$ is a member of one of these intervals. The function \textproc{EstimateGas} finds a local minimum for any non-reverting gas interval $[l_i,u_i]$ it reaches, that is, for which variable $g \in [l_i,u_i]$ during this function execution. Hence, if it misses the sub-interval where the (global) minimum gas limit is, this value is not found. So, it may yield either a local minimum or an undefined value. We use the function \lstinline{discontinuity} to illustrate this fact. We generate an \gls{evm} bytecode for this function such that its minimum gas limit is 21275. This function has two non-reverting intervals $i_1 = [21275,51223]$ and $i_2 = [20021255,g_\textit{top}]$. If we employ the binary search methodology, it misses the sub-interval $i_1$ and finds the local minimum for $i_2$, namely, it returns $\hat{g}_\textit{min} = 20021255$.

\begin{figure}[h]
\begin{lstlisting}
function discontinuity() public {
    if (30000 < gasleft() && gasleft() < 20000000) {
        revert();
    }
}    
\end{lstlisting}
\caption{Example of a function for which binary search does not find the minimum gas limit.}
\label{figure:enter-label}
\end{figure}

Typically, however, \glspl{sc} in Ethereum are naturally designed so that their executions have a single non-reverting gas interval $[g_\textit{min}, g_\textit{top}]$. Given that the binary search starts at $g_\textit{top}$ (reaching this interval), it proceeds to find $g_\textit{min}$.

In this paper, we are examining the hypothesis that one can reasonably accurately estimate the gas used and the minimum gas limit using these functions. Let us say that $T$ is a new transaction that a participant wants to send to the Ethereum blockchain, that the blockchain is currently at block $B^*$, and that $\sigma'$ is the state from which the transaction is executed when it finally gets added to a later block $B^\dagger$ and is processed by the blockchain. Our conjecture is that $\textproc{TraceCall}(\sigma_{B^*}, T)$ can be used to estimate $\Upsilon^g(\sigma', T)$, whereas $\textproc{EstimateGas}(\sigma_{B^*}, T)$ can be used to estimate $\textit{min}_g(\sigma',T)$, where $\sigma_{B^*}$ is the blockchain state right after the last transaction in $B^*$ is processed. Intuitively, in general, the precision of this estimation should degrade as the distance between $B^*$ and $B^\dagger$ increases; the more transactions between $B^*$ and $B^\dagger$, the more likely it is that some of them affects state $\sigma_{B^*}$ turning it into $\sigma'$ in a way that the execution of $T$ from $\sigma'$ is distinct from the one from $\sigma_{B^*}$, leading to a different gas used and minimum gas limit. We conjecture, however, that there is a class of \glspl{sc} whose executions would not be much (if at all) affected by changes in states. These contracts have a control flow that does not depend on blockchain state elements. Hence, their gas used and minimum gas limit would not change as the blockchain state evolves.

\section{Evaluation}\label{section:evaluation}

In the previous section, we have demonstrated that the minimum gas limit and the gas used are not equivalent, that is, there are cases where these two metrics diverge for the same transaction and blockchain state. Of course, in practice, it could still be that these cases where they diverge were only hardly exercised so this discrepancy would not be practically significant. In this section, however, we conduct an empirical study that demonstrates that for real transactions coming from a fragment of Ethereum's history, the difference between these two metrics is statistically significant. Intuitively, this result provides evidence that real transactions often fall into these cases where these two metrics diverge. Simply put, the previous section provides a qualitative analysis of the differences between the minimum gas limit and the gas used, whereas this section provides a quantitative one. Furthermore, we propose an estimation strategy (giving rise to an estimator) for each of these metrics which we test on the same fragment of Ethereum's history. The empirical studies that we conduct for these estimators suggest that they are very precise in predicting these metrics.

More precisely, the quantitative analysis of the differences between the minimum gas limit and the gas used and the evaluation of our estimators are carried out to answer the research questions outlined in this paper. We designed an experiment with three distinct stages, referred to as Experiments $E_{1}$, $E_{2}$, and $E_{3}$, each of which is used to answer the correspondingly numbered research question.

\subsection{Experimental Design}\label{subsection:experimental-design}

Our experiments use transactions collected from the Ethereum Mainnet's \textit{Bellatrix} fork, which starts at block 15,481,719 and ends at block 15,537,393, containing a total of 55,674 blocks. We selected this specific fork to achieve the best balance between the number of blocks available in our sample and the computational cost of processing and locally storing the transactions in a format compatible with our analysis. Additionally, this fork ensures compatibility with the latest updates regarding Ethereum's gas system. We collected, for each block, the first transaction, if it is an \gls{sc} function call and has been executed successfully (i.e. status code is 1). We gathered 38,174 transactions with this selection process.  The selection of only contract calls is somewhat uncontroversial. Simple transfers have a fixed minimum gas limit and gas used so there is no need for an estimator framework. We are excluding contract creation transactions from our analysis as they represent a very small percentage of transactions (0.03\% in our sample) and there are some implementation constraints with the client we use: it does not track contract initialisation code. The selection of first transactions was driven by implementation constraints. The execution layer client that we use does not support the calculation of minimum gas limit for intermediate block states; 
it only calculates this limit for the final state of a block. 
Thus, by using the first transaction of block $B$, we can use the final state of the previous block ($B^{-1}$) to try and have an accurate calculation. This calculation is still not exact; although the storage (i.e. state) of \glspl{sc} is the same at the end of block $B^{-1}$ and at the beginning of block $B$, the block context for the processing of $T$ as the last transaction of $B^{-1}$ is not the same as that when $T$ is the first transaction of $B$. For instance, the execution of 
a $T$ may use, in a distinctive way, the hash of the block it belongs to. In this case, the execution of $T$ in $B^{-1}$ and $B$ will be different. There are other elements of a block context that an execution can access and be influenced by. There are also transactions that are oblivious to this block context and their executions do not change whether they start from the final state of $B^{-1}$ or the starting state of $B$. Therefore, these experiments are based upon real transactions but they analyse executions that might be
slightly different than, but close to, the real executions of these transactions.

We conducted the experiments described in this section using a cluster with two Intel\textsuperscript{\textregistered} Xeon\textsuperscript{\textregistered} Gold 6338 processors, 2 TB of RAM, 22 TB SSD, and Ubuntu 22.04 LTS. In this cluster, we run an archive Ethereum node with Go Ethereum 1.10.21\footnote{https://geth.ethereum.org/}.

The Experiments $E_{1}$ and $E_{2}$ assess the precision of \textproc{EstimateGas} and \textproc{TraceCall} in, respectively, estimating the minimum gas limit and gas used of transactions. We compare the precision of these functions with that of the \gls{rgum} estimator proposed by Zarir et al.~\cite{Zarir:2021}. It estimates the gas used of a given transaction $T$ as the mean of the gas used of the last 10 transactions sent to the function called by $T$. We use this estimator and variations \gls{rgum-median}, \gls{rgum-max}, and \gls{rgum-min} that work similarly to \gls{rgum}, only changing the mean to the median, maximum, and minimum metrics, respectively. We also use \gls{rgum} (and its variations) to estimate the minimum gas limit because its authors suggest it can be used to find the gas limit for a transaction.

Intuitively, these experiments assess the precision of estimators in predicting the minimum gas limit and gas used values and how the evolution of the state being used in these predictions affects their accuracy. For each selected transaction $T$ of a block $B$, we estimate the minimum gas limit and gas used considering the state at the end of blocks $B^{-\Delta}$ for $\Delta \in \{1,6,11,21,101\}$. Considering that block
$B$ is the block at height $h$, $B^{-\Delta}$ is a previous block at height $h - \Delta$.
For example, $B^{-1}$ is the predecessor of $B$. We cannot conduct our experiment moving a  transaction $T$ forward (to blocks more recent than $B$), since $T$ could not be eliminated from $B$ and this would cause the duplication of the transaction, which is not allowed in Ethereum. Therefore, we execute the transaction in past blocks and investigate how the block gap $\Delta$ affects the estimation, namely, what is the error if one tries to use the minimum gas limit and gas used values at the end of block $B^{-\Delta}$ to predict the minimum gas limit at the end of $B^{-1}.$

The sort of \emph{age degradation for estimations} measured by this experiment is also useful to understand how predictions for the future should behave. Our measurements should shed light on how an estimation made on the current head block behaves as the chain evolves. This analysis is essential to understand whether these estimators can be useful in practice, namely, whether they can be used to estimate minimum gas limit and gas used by future transactions based on current information.

We quantified the precision of the evaluated estimators using the following metrics: \gls{ape} and $R^2$. These two metrics were calculated for each estimator and for every value of $\Delta$. The set of estimators evaluated and values of $\Delta$ are the independent variables and the estimated minimum gas limit and gas used are the dependent variables in Experiments $E_{1}$ and $E_{2}$, respectively.

The \gls{ape} \cite{Makridakis:1982} measures how much a given estimated value differs from the true value and is defined by Equation \ref{equation:ape}, where $y_{i}$ and $\hat{y}_{i}$ are the true value and the estimated value of observation $i$, respectively.

\begin{equation} \label{equation:ape}
    APE_{i} = \left| \frac{y_{i} - \hat{y}_{i}}{y_{i}} \right| \times 100
\end{equation}

The $R^2$ \cite{Draper:1998} evaluates the goodness of fit of the estimators. Equation \ref{equation:r2} presents the calculation of $R^2$. Similarly to Equation \ref{equation:ape}, $y_{i}$ and $\hat{y}_{i}$ represent the true value and the estimated value, respectively. In addition, $\bar{y}$ is the mean of the true values.

\begin{equation} \label{equation:r2}
    R^2 = 1 - \frac{\sum_i (y_{i} - \hat{y}_{i})^2}{\sum_i (y_{i} - \bar{y})^2}
\end{equation}

The values calculated by \textproc{EstimateGas} and \textproc{TraceCall} for $\Delta = 1$ are the true values for the minimum gas limit and gas used by the input transaction, respectively. So, there are no errors for \textproc{EstimateGas} and \textproc{TraceCall} at $\Delta = 1$, but, due to the state variation, typically there are errors for the other values of $\Delta$ and estimators. For each value of $\Delta$ and estimator, we compute the median, mean, and standard deviation of their \gls{ape}s, as well as the $R^2$ of their estimates.

For our analysis, we consider the following opcodes (EVM bytecode instructions) that are sensitive to information of the block context \cite{Wood:2014}: \ttt{BLOCKHASH}, \ttt{COINBASE}, \ttt{TIMESTAMP}, \ttt{NUMBER}, 
\ttt{GASLIMIT}, \ttt{BASEFEE}, \ttt{DIFFICULTY} \footnote{The opcode \ttt{DIFFICULTY} was replaced by \ttt{PREVRANDAO} following the Paris Fork \cite{Ethereumhist}, which implemented the Merge on the Ethereum blockchain.}. Moreover, the opcode \ttt{GAS}, which yields the remaining gas budget, suggests that the execution may be affected by the gas budget. We separate the collected transactions into two groups, according to how sensitive they should be to changes in the block context: Dataset $D_{1}$ (4,875 transactions) contains the ones whose executions do not involve any of the listed opcodes, whereas Dataset $D_{2}$ (33,299 transactions) contains the ones involving at least one of these opcodes. 
Hence, we can assess the ability of estimators in tackling these two classes separately. As we analyse both groups, this separation does not induce a bias in our experiments. Finally, this separation is also used to validate the sanity of the functions \textit{eth\_estimateGas()} and \textit{debug\_traceCall()} as correct implementations for \textproc{EstimateGas} and \textproc{TraceCall}. The transactions in Dataset $D_{1}$ must lead to the same execution, regardless of whether they are the first transaction of $B$ or the last one of $B^{-1}$; we indeed confirmed that.

In Experiment $E_{1}$, we defined the minimum gas limit estimators as the independent variable and the estimated minimum gas limit as the dependent variable. Similarly, in Experiment $E_{2}$, we defined the gas estimators as the independent variable and the estimated gas used as the dependent variable. We applied a randomized block design with five factors and two samples \cite{Wohlin:2012}\footnote{We waived the requirement for randomization in the application of the gas estimators for each sample, considering that the content of the transactions is not affected by their respective estimates.}, in which each group of estimators working at the same $\Delta$ level represents a factor, and the two samples are composed by the datasets $D_{1}$ and $D_{2}$. It is important to note that, except for the cases of \textproc{EstimateGas} with $\Delta = 1$ in Experiment $E_{1}$ and \textproc{TraceCall} with $\Delta = 1$ in Experiment $E_{2}$, there is no guarantee that the estimators will be able to predict the gas for all transactions in datasets $D_{1}$ and $D_{2}$. To ensure consistency across the experiments, we included in our results only the transactions in which all gas estimators of the same factor returned a valid estimate.

Following the same strategy adopted by the authors of \gls{rgum} \cite{Zarir:2021}, we chose to rank the precision of the estimators based on their medians. For each dataset, we selected the \gls{rgum} version with the lowest median \gls{ape} values to compare with \textproc{EstimateGas} in Experiment $E_{1}$, and with \textproc{TraceCall} in Experiment $E_{2}$. After selecting the estimators, we present boxplots of the distributions of their \glspl{ape} to examine how their medians relate to the other quartiles and outliers.

To verify whether there are statistical differences between the results obtained by the estimators, we compared the \glspl{ape} of their estimates in both analysed datasets by applying the \textit{Kruskal-Wallis} test \cite{Kruskal:1952}, using a 95\% significance level. This nonparametric test is used to determine whether there are significant differences between three or more samples of independent data, without requiring assumptions such as the normality of the evaluated data.

We define the following hypotheses for Experiment $E_{1}$:

\begin{itemize}

    \item Null hypothesis for Experiment $E_{1}$ ($H_{0, E_{1}}$): There is no statistical difference between the distributions of the \glspl{ape} of the \textproc{EstimateGas} and the other estimators for all the considered values of $\Delta$s.

    \item Alternative hypothesis for Experiment $E_{1}$ ($H_{1, E_{1}}$): The distribution of the \glspl{ape} of the \textproc{EstimateGas} is statistically different from the distribution of the \glspl{ape} of at least one of the other estimators for all the considered values of $\Delta$s.
    
\end{itemize}

For Experiment $E_{2}$, the hypotheses are defined as follows:

\begin{itemize}

    \item Null hypothesis for Experiment $E_{2}$ ($H_{0, E_{2}}$): There is no statistical difference between the distributions of the \glspl{ape} of the \textproc{TraceCall} and the other estimators for for all the considered values of $\Delta$s.

    \item Alternative hypothesis for Experiment $E_{2}$ ($H_{1, E_{2}}$): The distribution of the \glspl{ape} of the \textproc{TraceCall} is statistically different from the distribution of the \glspl{ape} of at least one of the other estimators for all the considered values of $\Delta$s.
    
\end{itemize}

Since the \textit{Kruskal-Wallis} test only indicates whether there are statistical differences between the samples analysed, without detailing how the results differ statistically from each other, we applied the \textit{Conover} post hoc test \cite{Conover:1999} to identify in which contexts each estimator was, or was not, statistically superior to the others in our experiments.

For each value of $\Delta$ in both experiments, the application of the \textit{Kruskal-Wallis} test, together with the post hoc \textit{Conover} test, generates a $10 \times 10$ matrix of results. Following the same strategy adopted in the presentation of the boxplots, we perform statistical comparisons between the \gls{ape} distributions (for varying $\Delta$s) obtained by the \textit{RGUM} variation with the least median value and those obtained by \textproc{EstimateGas} in Experiment 1, and by \textproc{TraceCall} in Experiment 2. For better visualisation, we presented the results of these comparisons using heatmaps.

For experiment $E_{3}$, we analyse the difference between the gas used and the minimum gas limit of transactions. For this, we use the \gls{ape} calculation in Equation~\ref{equation:ape} where $y_i$ is the gas used and $\hat{y}_i$ is the minimum gas limit for a transaction; we also calculate $R^2$ using these values. Therefore, we consider Ethereum transaction parameters as the independent variables and the respective gas used as the dependent variable. 
In this context, high \gls{ape} values suggest greater discrepancies between these metrics. On the other hand, $R^{2}$ values closer to 1 indicate a strong linear correlation between the minimum gas limit and gas used values.

We applied a randomized complete block design with two samples \cite{Wohlin:2012} in Experiment $E_{3}$, similarly to what we did in Experiments $E_{1}$ and $E_{2}$, where Dataset $D_{1}$ corresponds to the first sample and Dataset $D_{2}$ corresponds to the second sample.
For each of the two datasets analysed, we applied the two-sample \gls{ks} test \cite{Smirnov:1948} with a 95\% significance level to determine whether there are significant differences between the minimum gas limit and gas used distributions of the transactions, presenting the results with \gls{ks} plots. This nonparametric test aims to compare two samples of continuous values to verify whether they come from the same distribution. Capable of identifying differences between two distributions that go beyond measures of location or dispersion, this test is widely used when one wants to verify whether two distributions have the same shape.

The hypotheses that we defined for Experiment $E_{3}$ are as follows:

\begin{itemize}

    \item Null hypothesis $E_{3}$ ($H_{0, E_{3}}$): There are no statistical differences between the distributions of minimum gas limit and gas used by the analysed transactions. 

    \item Alternative hypothesis $E_{3}$ ($H_{1, E_{3}}$): The distributions of minimum gas limit and gas used by the analysed transactions are statistically different. 
    
\end{itemize}

\subsection{Minimum Gas Limit Experiment ($E_1$)} \label{subsection:minimum-gas-limit-experiment}

In this subsection, we present the results of Experiment $E_{1}$, following the protocol defined in \ref{subsection:experimental-design}. We detail the outcomes of the \textproc{EstimateGas} applied to Datasets $D_{1}$ and $D_{2}$, as described below.

For each value of $\Delta$ and estimator, we compute the median, mean, and standard deviation of their \gls{ape}s, as well as the $R^2$ of their estimates; the results for both datasets are presented in Tables \ref{table:minimum-gas-limit-median}, \ref{table:minimum-gas-limit-mean}, \ref{table:minimum-gas-limit-r2}. As expected, in most cases, the precision of the estimators decreases as $\Delta$ increases; 
as we consider earlier configurations of the blockchain,
its state is more likely to have changed, and more substantially so. Moreover, the precision of all estimators consistently decreases as they move from Dataset $D_{1}$ to Dataset $D_{2}$; the fact that transactions in Dataset 2 have opcodes that are more sensitive to blockchain state changes than the ones in Dataset 1 is a reasonable justification for this pattern.

\begin{table}[!ht]
\centering
\begin{tabular}{|c|c|c|c|c|c|c|}
\hline
\multicolumn{7}{|c|}{Dataset 1} \\
\hline
  $\Delta$&
  \textbf{\#Txs} &
  \textproc{EstimateGas} &
  \textit{RGUM} &
  \textit{RGUM}$_\textit{median}$ &
  \textit{RGUM}$_\textit{max}$ &
  \textit{RGUM}$_\textit{min}$ \\ \hline
1     & 4380                   & 0      & 13.62      & 10.43        & 2.22      & 13.55     \\ \hline
6     & 4145                   & 0      & 13.76      & 11.27        & 2.22      & 13.68     \\ \hline
11    & 3971                   & 0      & 13.9       & 12.38        & 0.65      & 13.78     \\ \hline
21    & 3723                   & 0      & 14.29      & 13.35        & 0.62      & 13.83     \\ \hline
101   & 3133                   & 0      & 14.76      & 13.52        & 0.18      & 15.61     \\ \hline
\multicolumn{7}{c}{} \\
\hline
\multicolumn{7}{|c|}{Dataset 2} \\
\hline
  $\Delta$&
  \textbf{\#Txs} &
  \textproc{EstimateGas} &
  \textit{RGUM} &
  \textit{RGUM}$_\textit{median}$ &
  \textit{RGUM}$_\textit{max}$ &
  \textit{RGUM}$_\textit{min}$ \\ \hline
1     & 22970                  & 0      & 19.04      & 15.34        & 22.19     & 45.0      \\ \hline
6     & 17656                  & 0      & 19.31      & 15.43        & 22.4      & 49.13     \\ \hline
11    & 16092                  & 0      & 19.23      & 15.23        & 22.57     & 49.1      \\ \hline
21    & 14798                  & 0.01     & 19.52      & 15.62        & 23.23     & 50.5      \\ \hline
101   & 12220                  & 0.55     & 20.12      & 16.2         & 24.15     & 56.43     \\ \hline
\end{tabular}
\caption{Comparison of median \glspl{ape} for minimum gas limit estimators.}
\label{table:minimum-gas-limit-median}
\end{table}

According to Table \ref{table:minimum-gas-limit-median}, \textproc{EstimateGas} outperformed the other estimators, presenting a median \gls{ape} equal to 0
in all cases, except for its versions with $\Delta = 21$ and $\Delta = 101$, when used to estimate the minimum gas limit of transactions belonging to Dataset $D_{2}$, where the results were 0.01\% and 0.55\%, respectively. In Dataset $D_{1}$, \gls{rgum-max} was the \gls{rgum} version that exhibited the best results, with median \glspl{ape} ranging from 0.18\% to 2.22\%. It is important to note that, unlike the other \gls{rgum} versions, \gls{rgum-max} showed a reduction in its median \gls{ape} as delta values increased. In Dataset $D_{2}$, the \gls{rgum} version that presented the lowest median \glspl{ape} was \gls{rgum-median}, following the pattern of increasing error as the $\Delta$ value increased, with values ranging from 15.34\% to 16.2\%.

\begin{table}[h]
\centering
\begin{tabular}{|c|c|c|c|c|c|c|}
\hline
\multicolumn{7}{|c|}{Dataset 1} \\
\hline
  $\Delta$ &
  \textbf{\#Txs} &
  \textproc{EstimateGas} &
  \textit{RGUM} &
  \textit{RGUM}$_\textit{median}$ &
  \textit{RGUM}$_\textit{max}$ &
  \textit{RGUM}$_\textit{min}$ \\ \hline
1   & 4380 & 0 (0)    & 15.23 (16.03)  & 15.27 (17.92)  & 21.02 (51.04)  & 18.59 (17.3)  \\ \hline
6   & 4145 & 0 (5.12) & 15.38 (16.13)  & 15.61 (18.35)  & 20.38 (42.69)  & 19.15 (17.72) \\ \hline
11  & 3971 & 0.81 (6.4)  & 17.16 (107.66) & 17.25 (107.88) & 23.56 (217.46) & 19.52 (17.71) \\ \hline
21  & 3723 & 1.07 (7.34) & 15.93 (18.51)  & 16.29 (19.21)  & 20.51 (44.03)  & 19.88 (17.89) \\ \hline
101 & 3133 & 2.04 (9.63) & 16.42 (17.95)  & 16.34 (18.73)  & 20.78 (52.34)  & 21.51 (16.97) \\ \hline
\multicolumn{7}{c}{} \\
\hline
\multicolumn{7}{|c|}{Dataset 2} \\
\hline
  $\Delta$ &
  \textbf{\#Txs} &
  \textproc{EstimateGas} &
  \textit{RGUM} &
  \textit{RGUM}$_\textit{median}$ &
  \textit{RGUM}$_\textit{max}$ &
  \textit{RGUM}$_\textit{min}$ \\ \hline
1   & 22970 & 0 (0)     & 34.94 (72.4)  & 33.46 (72.62) & 55.95 (115.85) & 52.57 (60.6)  \\ \hline
6   & 17656 & 6.99 (48.95) & 37.35 (74.85) & 35.69 (74.72) & 58.53 (121.95) & 54.91 (60.37) \\ \hline
11  & 16092 & 7.65 (50.02) & 38.37 (77.65) & 36.79 (77.11) & 59.92 (124.86) & 55.64 (62.39) \\ \hline
21  & 14798 & 8.5 (53.15) & 39.5 (79.47) & 38.01 (79.15) & 62.04 (128.49) & 56.23 (63.3)  \\ \hline
101 & 12220 & 9.3 (38.72) & 41.71 (85.22) & 39.89 (83.62) & 65.32 (141.93) & 58.16 (64.66) \\ \hline
\end{tabular}
\caption{Comparison of the mean (with standard deviation in parentheses) for \glspl{ape} of minimum gas limit estimations.}
\label{table:minimum-gas-limit-mean}
\end{table}

Listing the means and standard deviations of the \glspl{ape}, Table \ref{table:minimum-gas-limit-mean} once again shows that \textproc{EstimateGas} outperformed all \gls{rgum} versions in both datasets. In Dataset $D_{1}$, the mean \gls{ape} of \textproc{EstimateGas} ranged from 0\% to 2.04\%. In Dataset $D_{2}$, the values ranged from 0\% to 9.3\%. In both cases, the mean \glspl{ape} of \textproc{EstimateGas} increased consistently as the delta values increased. Unlike what was seen in Table \ref{table:minimum-gas-limit-median}, the \gls{rgum} version that obtained the best results in Dataset $D_{1}$ was the original \gls{rgum}, with a mean \glspl{ape} ranging from 15.23\% to 17.16\%. It is interesting to note that all \gls{rgum} versions, except for \gls{rgum-min}, recorded their highest mean APE when $\Delta = 11$, also reporting much higher standard deviation values than in the other cases. We investigated the reason behind this unusual behavior and found that the mean \glspl{ape} of these three \gls{rgum} versions were particularly affected by a single transaction. This finding reinforces the need to use the median as a reference metric for statistical comparisons between estimators, as it is less sensitive to the presence of outliers than the mean. In Dataset $D_{2}$, the lowest mean \glspl{ape} were achieved by \gls{rgum-median}, with values that increased in line with $\Delta$, ranging from 33.46\% at $\Delta = 1$ to 39.89\% at $\Delta = 101$.

\begin{table}[!hb]
\centering
\begin{tabular}{|c|c|c|c|c|c|c|}
\hline
\multicolumn{7}{|c|}{Dataset 1} \\
\hline
  $\Delta$&
  \textbf{\#Txs} &
  \textproc{EstimateGas} &
  \textit{RGUM} &
  \textit{RGUM}$_\textit{median}$ &
  \textit{RGUM}$_\textit{max}$ &
  \textit{RGUM}$_\textit{min}$ \\ \hline
1     & 4380                   & 1      & 0.9455     & 0.9344       & 0.8427    & 0.904     \\ \hline
6     & 4145                   & 0.9989   & 0.9515     & 0.944        & 0.8844    & 0.9179    \\ \hline
11    & 3971                   & 0.9984   & 0.5702     & 0.5678       & 0.2532    & 0.9231    \\ \hline
21    & 3723                   & 0.9981   & 0.9533     & 0.9484       & 0.8832    & 0.925     \\ \hline
101   & 3133                   & 0.9967   & 0.9553     & 0.9537       & 0.8426    & 0.9251    \\ \hline
\multicolumn{7}{c}{} \\
\hline
\multicolumn{7}{|c|}{Dataset 2} \\
\hline
  $\Delta$&
  \textbf{\#Txs} &
  \textproc{EstimateGas} &
  \textit{RGUM} &
  \textit{RGUM}$_\textit{median}$ &
  \textit{RGUM}$_\textit{max}$ &
  \textit{RGUM}$_\textit{min}$ \\ \hline
1     & 22970                  & 1      & -0.3185    & -0.5275      & 0.2692    & -2.451    \\ \hline
6     & 17656                  & 0.9641   & -0.3985    & -0.6833      & 0.3016    & -3.1857   \\ \hline
11    & 16092                  & 0.9687   & -0.3346    & -0.6116      & 0.3407    & -3.1592   \\ \hline
21    & 14798                  & 0.9642   & -0.3427    & -0.6444      & 0.3342    & -3.2453   \\ \hline
101   & 12220                  & 0.9613   & -0.439     & -0.7964      & 0.2769    & -3.0323   \\ \hline
\end{tabular}
\caption{Comparison of $R^{2}$ for minimum gas limit estimators.}
\label{table:minimum-gas-limit-r2}
\end{table}

Table \ref{table:minimum-gas-limit-r2} lists the results obtained by the estimators according to the $R^{2}$ metric. Once again, \textproc{EstimateGas} showed the best results, with $R^{2}$ values consistently decreasing as the $\Delta$ value increased, ranging from 1 to 0.9967 in Dataset $D_{1}$, and from 1 to 0.9613 in Dataset $D_{2}$. In Dataset $D_{1}$, consistent with the results presented in Table \ref{table:minimum-gas-limit-mean}, the original \gls{rgum}, along with \gls{rgum-median} and \gls{rgum-max}, reported their worst precisions when $\Delta = 11$, with $R^{2}$ values considerably lower than those in the other cases. \gls{rgum-min} achieved more balanced results, with $R^{2}$ ranging from 0.904 to 0.9251. It should be noted that, unlike \textproc{EstimateGas}, the $R^{2}$ values of \gls{rgum-min} increases as  $\Delta$ increases. \gls{rgum-max} was the only version of \gls{rgum} that obtained positive $R^{2}$ values in Dataset $D_{2}$, with values ranging from 0.2692 to 0.3407. These results are interesting because they indicate that, although \gls{rgum-max} did not present the lowest median and mean \glspl{ape}, it was the version of \gls{rgum} that demonstrated the highest level of linear relationship and, consequently, the highest level of explainability between its results and the true values of the minimum gas limit.

\begin{figure*}[ht]
    \centering
    \includegraphics[width=370px]{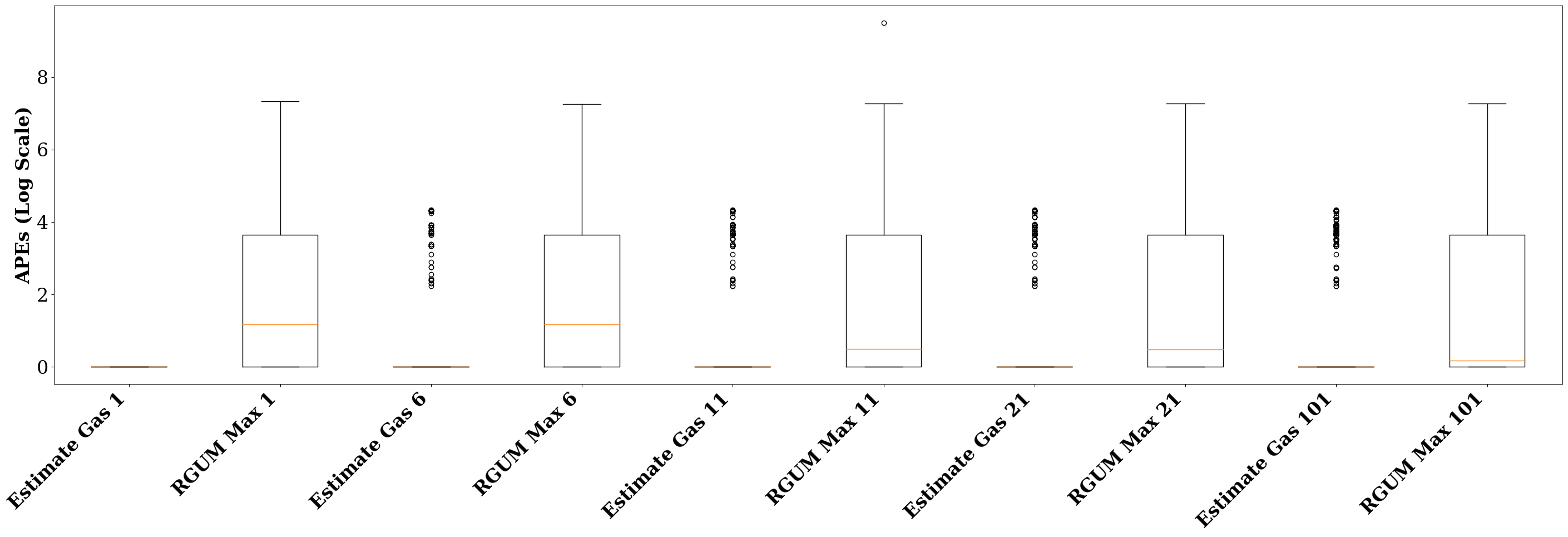}
    \caption{Boxplot of \glspl{ape} for minimum gas limit estimators for Dataset $D_{1}$.}
    \label{image:boxplot-minimum-gas-limit-good-opcodes}
\end{figure*}

\begin{figure*}[ht]
    \centering
    \includegraphics[width=370px]{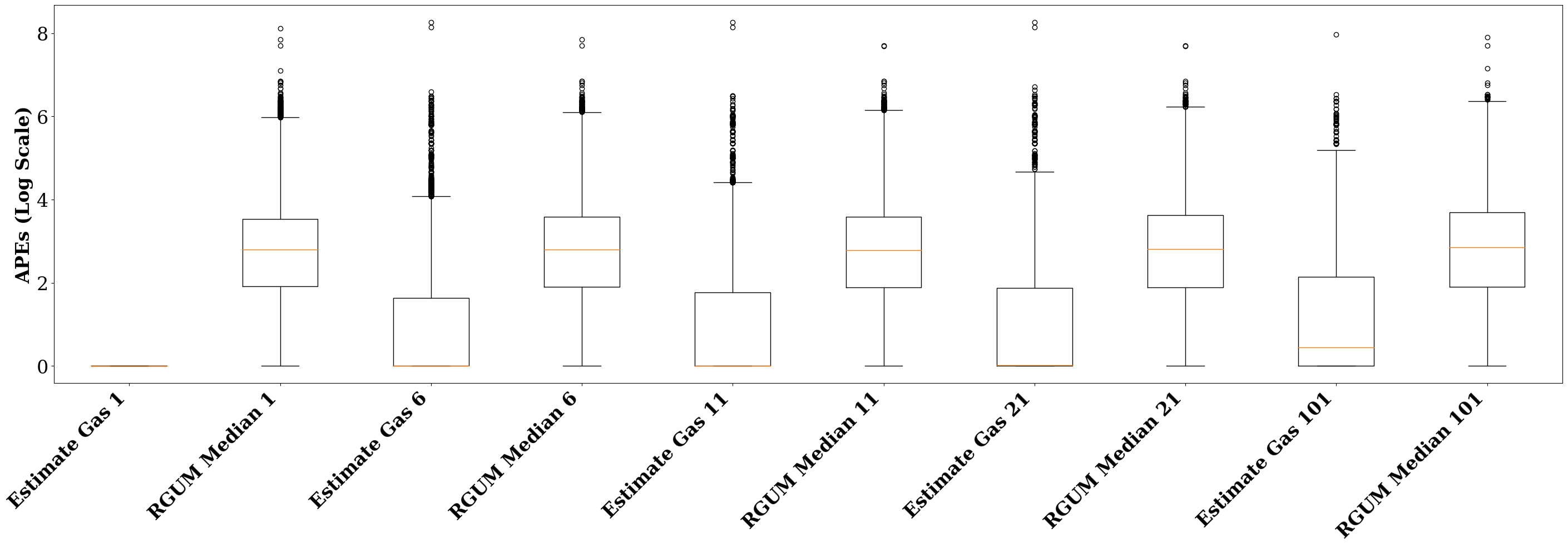}
    \caption{Boxplot of \glspl{ape} for minimum gas limit estimators for Dataset $D_{2}$.}
    \label{image:boxplot-minimum-gas-limit-bad-opcodes}
\end{figure*}

Figures \ref{image:boxplot-minimum-gas-limit-good-opcodes} and \ref{image:boxplot-minimum-gas-limit-bad-opcodes}  present boxplots showing the distributions of the \glspl{ape} for \textproc{EstimateGas} and the \gls{rgum} versions that had the lowest median \glspl{ape} in the two analysed datasets. In Dataset $D_{1}$, the medians, first quartiles, and third quartiles of \textproc{EstimateGas}’s \glspl{ape} for all $\Delta$ levels were equal to 0. This shows that even when $\Delta = 101$, \textproc{EstimateGas} estimated the exact minimum gas limit value for at least 50\% of the transactions in this dataset. Furthermore, for all $\Delta$ values, \textproc{EstimateGas}’s estimation errors in Dataset $D_{1}$, represented by the outliers associated with this estimator in Figure \ref{image:boxplot-minimum-gas-limit-good-opcodes}, were lower than the upper whiskers of the boxplots for all \gls{rgum-max} cases, reinforcing the higher precision of \textproc{EstimateGas}. Also in Figure \ref{image:boxplot-minimum-gas-limit-good-opcodes}, it is interesting to note that there is a single outlier in the \gls{rgum-max} boxplot with $\Delta = 11$, which confirms what we described about how the mean \glspl{ape} of three \gls{rgum} versions were particularly affected by a single transaction for this $\Delta$ value. The median and first quartile values of the \textproc{EstimateGas} \gls{ape} distributions when estimating the minimum gas limit for transactions in Dataset $D_{2}$ were 0 at all $\Delta$ levels, except for $\Delta = 21$ and $\Delta = 101$. For these $\Delta$ values, the \textproc{EstimateGas} \gls{ape} medians were 0.01 and 0.55, respectively, as listed in Table \ref{table:minimum-gas-limit-mean}. These values demonstrate that, despite its precision being consistently affected by the existence of block state-related opcodes, \textproc{EstimateGas} was able to predict the exact minimum gas limit for at least 25\% of transactions in Dataset $D_{2}$ when $\Delta < 101$. In addition, Figure \ref{image:boxplot-minimum-gas-limit-bad-opcodes} shows that the third quartile of all \textproc{EstimateGas} boxplots was below the median \gls{ape} of \gls{rgum-median} at all $\Delta$ levels, indicating that \textproc{EstimateGas} consistently maintained lower \gls{ape} values than those obtained by \gls{rgum-median}.

\begin{figure*}[ht]
    \centering
    \includegraphics[width=370px]{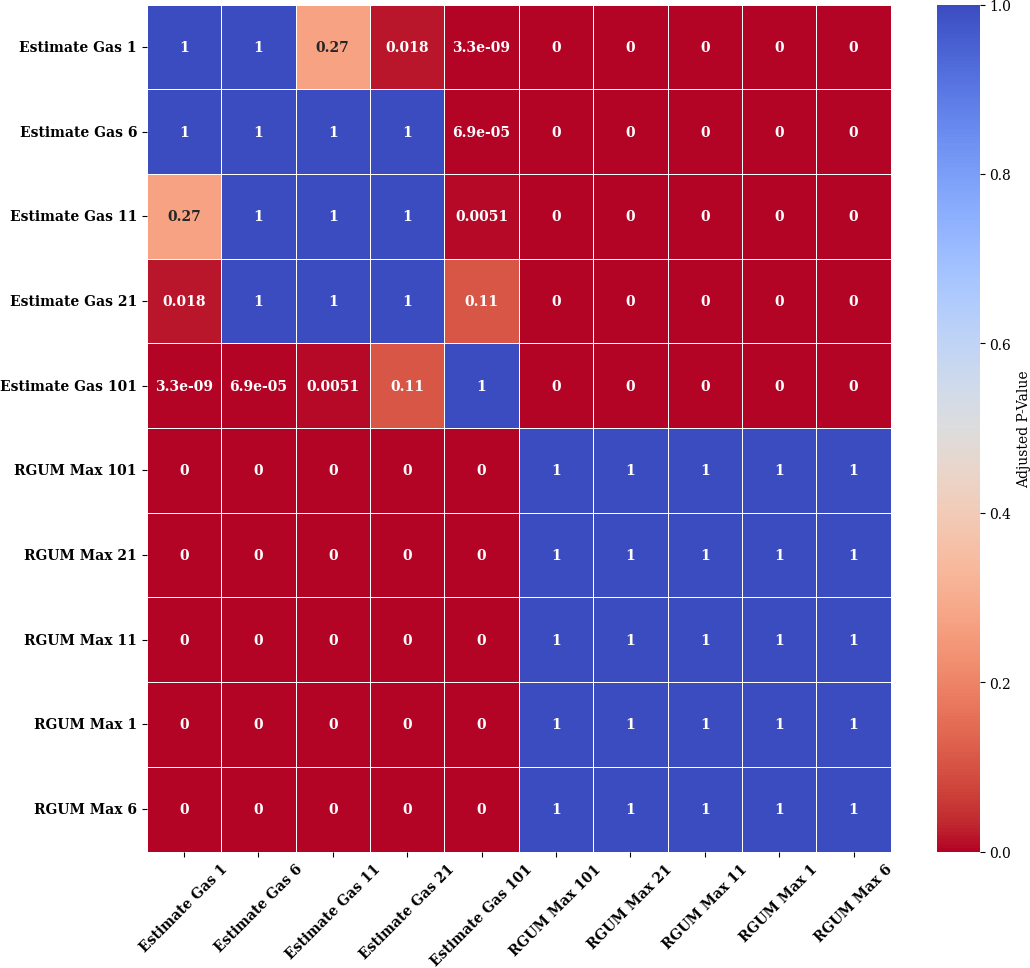}
    \caption{Heatmap of \glspl{ape} for minimum gas limit estimators for Dataset $D_{1}$.}
    \label{image:heatmap-minimum-gas-limit-good-opcodes}
\end{figure*}

\begin{figure*}[ht]
    \centering
    \includegraphics[width=370px]{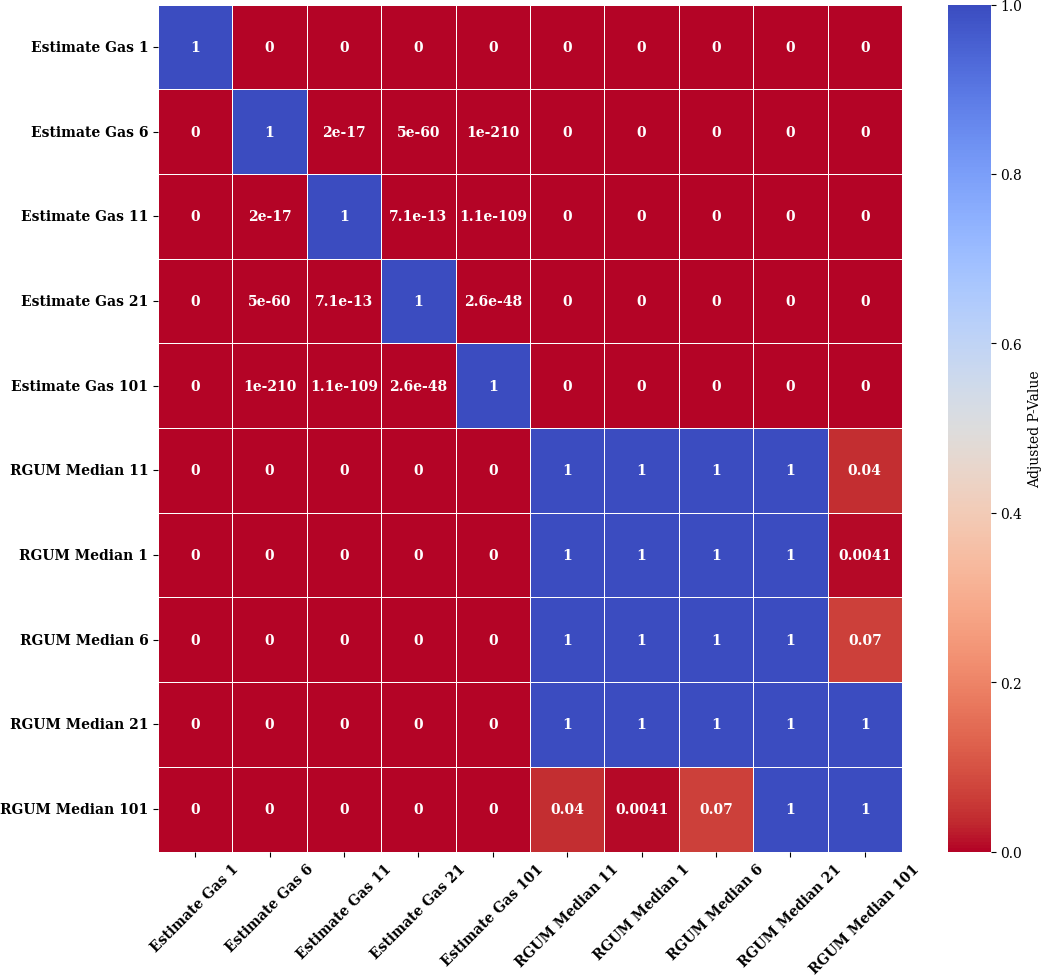}
    \caption{Heatmap of \glspl{ape} for minimum gas limit estimators for Dataset $D_{2}$.}
    \label{image:heatmap-minimum-gas-limit-bad-opcodes}
\end{figure*}

Figures \ref{image:heatmap-minimum-gas-limit-good-opcodes} and \ref{image:heatmap-minimum-gas-limit-bad-opcodes} present the heatmaps with the results of the statistical comparison between the \gls{ape} distributions shown in Figures \ref{image:boxplot-minimum-gas-limit-good-opcodes} and \ref{image:boxplot-minimum-gas-limit-bad-opcodes}, respectively, using the \textit{Kruskal-Wallis} test with the \textit{Conover} post hoc test. In both figures, it can be observed that the precision of \textproc{EstimateGas} in both datasets, when $\Delta = 101$, was statistically superior to the best cases of \gls{rgum-max} in Dataset $D_{1}$ and \gls{rgum-median} in Dataset $D_{2}$, which occurred when $\Delta = 101$ and $\Delta = 11$, respectively. In Dataset $D_{1}$, the statistical test reported no significant difference between the precision of \textproc{EstimateGas} when $\Delta = 1$ (which represents the perfect estimator), $\Delta = 6$, and $\Delta = 11$. Furthermore, the \textproc{p-value} for the comparison between the distributions of the \textproc{EstimateGas} \glspl{ape} with $\Delta = 1$ and $\Delta = 21$ was 0.018, which does not indicate a statistical difference between the two distributions at a level of significance 1\%. For this dataset, \textproc{EstimateGas} shown a statistically significant precision drop, at any significance level, only with $\Delta = 101$. As expected, as the $\Delta$ value increased in Dataset $D_{2}$, the precision of Estimate Gas decreased significantly. In this dataset, the statistical test reported a p-value close to 0 when comparing the APE distributions of \textproc{EstimateGas} at $\Delta = 1$ and $\Delta = 6$, reinforcing how block-state-dependent opcodes affect the minimum gas limit of transactions as the states of the block change. Based on these results, we reject the null hypothesis for Experiment $E_{1}$ ($H_{0, E_{1}}$).

\subsection{Gas Used Experiment ($E_2$)} \label{subsection:gas-used-experiment}

The experiment described in this subsection is very similar to that presented in Section \ref{subsection:minimum-gas-limit-experiment}. Here, we present the results of Experiment $E_{2}$, following the protocol defined in \ref{subsection:experimental-design}. We detail the outcomes of the \textproc{TraceCall} applied to  Datasets $D_{1}$ and $D_{2}$ as described in the following.

As in Section \ref{subsection:minimum-gas-limit-experiment}, we calculated the median, mean, and standard deviation of the \glspl{ape} of the estimators, as well as the $R^{2}$ of their estimates, for each $\Delta$ value. The results are presented in Tables \ref{table:gas-usage-median}, \ref{table:gas-usage-mean}, and \ref{table:gas-usage-r2}. Once again, the precision of the estimators decreased as the $\Delta$ values increased in most cases, and the estimates obtained in Dataset $D_{2}$ were less accurate than those in Dataset $D_{1}$.

Considering the three metrics used in Experiment $E_{2}$, we can see that when estimating the gas used by the transactions in Datasets $D_{1}$ and $D_{2}$, \textproc{TraceCall} performed at the same level as \textproc{EstimateGas} did when estimating minimum gas limit. Except for its worst case, in Dataset $D_{2}$ when $\Delta = 101$ under the $R^{2}$ metric, \textproc{TraceCall} estimated gas used with greater precision than all versions of \gls{rgum}, regardless of the $\Delta$ values considered.

\begin{table}[!ht]
\centering
\begin{tabular}{|c|c|c|c|c|c|c|}
\hline
\multicolumn{7}{|c|}{Dataset 1} \\
\hline
  $\Delta$&
  \textbf{\#Txs} &
  \textproc{TraceCall} &
  \textit{RGUM} &
  \textit{RGUM}$_\textit{median}$ &
  \textit{RGUM}$_\textit{max}$ &
  \textit{RGUM}$_\textit{min}$ \\ \hline
1     & 4380                   & 0   & 13.89      & 7.93         & 1.24      & 10.44     \\ \hline
6     & 4145                   & 0   & 13.98      & 9.26         & 0.63      & 10.44     \\ \hline
11    & 3971                   & 0   & 14.32      & 10.38        & 0.12      & 10.46     \\ \hline
21    & 3723                   & 0  & 14.91      & 11.62        & 0.18      & 11.59     \\ \hline
101   & 3133                   & 0   & 15.04      & 11.62        & 0.06      & 13.84     \\ \hline
\multicolumn{7}{c}{} \\
\hline
\multicolumn{7}{|c|}{Dataset 2} \\
\hline
  $\Delta$&
  \textbf{\#Txs} &
  \textproc{TraceCall} &
  \textit{RGUM} &
  \textit{RGUM}$_\textit{median}$ &
  \textit{RGUM}$_\textit{max}$ &
  \textit{RGUM}$_\textit{min}$ \\ \hline
1     & 22970                  & 0   & 17.3       & 12.8         & 25.75     & 41.45     \\ \hline
6     & 17656                  & 0   & 17.69      & 12.54        & 25.96     & 43.34     \\ \hline
11    & 16092                  & 0   & 17.7       & 12.38        & 26.12     & 43.5      \\ \hline
21    & 14798                  & 0.02  & 18.09      & 12.9         & 26.89     & 44.41     \\ \hline
101   & 12220                  & 0.8   & 18.7       & 13.56        & 27.42     & 48.97     \\ \hline
\end{tabular}
\caption{Comparison of median \glspl{ape} for gas used estimators.}
\label{table:gas-usage-median}
\end{table}

Table \ref{table:gas-usage-median} presents the medians of the \glspl{ape} obtained by the estimators when predicting gas used. It shows that \textproc{TraceCall} achieved a median \gls{ape} of 0 for all $\Delta$ values in Dataset $D_{1}$. However, in Dataset $D_{2}$, it failed to achieve this result when $\Delta = 21$ and $\Delta = 101$, where its median \glspl{ape} were 0.02 and 0.8, respectively. Repeating the precision level observed in Experiment $E_{1}$, \gls{rgum-max} and \gls{rgum-median} obtained the lowest median \glspl{ape} for Dataset $D_{1}$ and Dataset $D_{2}$, respectively. In Dataset $D_{1}$, the median \gls{ape} of \gls{rgum-max} ranged between 0.06 and 1.24, with the best result when $\Delta = 101$ and the worst when $\Delta = 1$. In Dataset $D_{2}$, \gls{rgum-median} achieved median \glspl{ape} between 12.38 and 13.56.

\begin{table}[!ht]
\centering
\begin{tabular}{|c|c|c|c|c|c|c|}
\hline
\multicolumn{7}{|c|}{Dataset 1} \\
\hline
  $\Delta$&
  \textbf{\#Txs} &
  \textproc{TraceCall} &
  \textit{RGUM} &
  \textit{RGUM}$_\textit{median}$ &
  \textit{RGUM}$_\textit{max}$ &
  \textit{RGUM}$_\textit{min}$ \\ \hline
1   & 4380 & 0 (0)    & 16.4 (18.35)  & 15.93 (20.32)  & 23.98 (54.75)  & 16.95 (18.69) \\ \hline
6   & 4145 & 0.51 (5.18) & 16.61 (18.47)  & 16.35 (20.78)  & 23.33 (46.17)  & 17.66 (19.09) \\ \hline
11  & 3971 & 0.84 (6.5)  & 18.50 (118.93) & 18.17 (119.17) & 26.7 (239.63) & 18.03 (19.11) \\ \hline
21  & 3723 & 1.11 (7.47) & 17.1 (20.76)  & 17.11 (21.51)  & 23.12 (47.41)  & 18.61 (19.21) \\ \hline
101 & 3133 & 2.14 (9.87) & 17.16 (19.86)  & 16.69 (20.68)  & 22.86 (54.82)  & 20.47 (18.06) \\ \hline
\multicolumn{7}{c}{} \\
\hline
\multicolumn{7}{|c|}{Dataset 2} \\
\hline
  $\Delta$&
  \textbf{\#Txs} &
  \textproc{TraceCall} &
  \textit{RGUM} &
  \textit{RGUM}$_\textit{median}$ &
  \textit{RGUM}$_\textit{max}$ &
  \textit{RGUM}$_\textit{min}$ \\ \hline
1   & 22970 & 0 (0)       & 24.98 (43)    & 23.01 (44.05) & 49.02 (94.5)   & 43.06 (38.88) \\ \hline
6   & 17656 & 6.25 (37.69)   & 25.36 (37.21) & 23.04 (38.02) & 50.18 (95.92)  & 43.22 (30.83) \\ \hline
11  & 16092 & 6.9 (38.99)   & 25.32 (37.66) & 23.02 (37.6)  & 50.1 (96.25)  & 43.24 (30.61) \\ \hline
21  & 14798 & 7.64 (40.58)   & 25.96 (39.05) & 23.73 (39.27) & 51.6 (99.72)  & 43.47 (30.96) \\ \hline
101 & 12220 & 10.83 (217.99) & 26.91 (45.48) & 24.68 (44.43) & 52.93 (112.48) & 44.66 (30.51) \\ \hline
\end{tabular}
\caption{Comparison of mean \glspl{ape} for gas used estimators.}
\label{table:gas-usage-mean}
\end{table}

In Table \ref{table:gas-usage-mean}, which presents the means and standard deviations of the \glspl{ape} for the gas used estimates, the mean \glspl{ape} of \textproc{TraceCall} ranged from 0 to 2.14 in Dataset $D_{1}$ and from 0 to 10.83 in Dataset $D_{2}$. Here, \gls{rgum-median} outperformed the other versions of \gls{rgum} in both datasets and for all $\Delta$ values, except when $\Delta = 21$ in Dataset $D_{1}$, where the mean \gls{ape} of the original \gls{rgum} was 17.1 and that of \gls{rgum-median} was 17.11. Also in Table \ref{table:gas-usage-mean}, it should be noted that, following the same pattern observed in Table \ref{table:minimum-gas-limit-mean}, the Original \gls{rgum}, \gls{rgum-median}, and \gls{rgum-max} obtained the worst mean \glspl{ape} in Dataset $D_{1}$ when $\Delta = 11$, where the standard deviations of their \glspl{ape} were also considerably higher than in other cases. This behavior is due to the same reason as described earlier when we presented the results of Experiment $E_{1}$.

\begin{table}[!ht]
\centering
\begin{tabular}{|c|c|c|c|c|c|c|}
\hline
\multicolumn{7}{|c|}{Dataset 1} \\
\hline
  $\Delta$&
  \textbf{\#Txs} &
  \textproc{TraceCall} &
  \textit{RGUM} &
  \textit{RGUM}$_\textit{median}$ &
  \textit{RGUM}$_\textit{max}$ &
  \textit{RGUM}$_\textit{min}$ \\ \hline
1     & 4380                   & 1    & 0.9446     & 0.9338       & 0.8366    & 0.9076    \\ \hline
6     & 4145                   & 0.9989 & 0.9505     & 0.9432       & 0.8783    & 0.9212    \\ \hline
11    & 3971                   & 0.9984 & 0.5678     & 0.5656       & 0.2506    & 0.9249    \\ \hline
21    & 3723                   & 0.9981 & 0.9513     & 0.9465       & 0.8779    & 0.9264    \\ \hline
101   & 3133                   & 0.9966 & 0.9543     & 0.9527       & 0.8392    & 0.9268    \\ \hline
\multicolumn{7}{c}{} \\
\hline
\multicolumn{7}{|c|}{Dataset 2} \\
\hline
  $\Delta$&
  \textbf{\#Txs} &
  \textproc{TraceCall} &
  \textit{RGUM} &
  \textit{RGUM}$_\textit{median}$ &
  \textit{RGUM}$_\textit{max}$ &
  \textit{RGUM}$_\textit{min}$ \\ \hline
1     & 22970                  & 1    & 0.0231     & -0.1332      & 0.2817    & -1.5498   \\ \hline
6     & 17656                  & 0.9687 & 0.0488     & -0.1404      & 0.3202    & -1.9257   \\ \hline
11    & 16092                  & 0.9662 & 0.0831     & -0.102       & 0.3557    & -1.9315   \\ \hline
21    & 14798                  & 0.9602 & 0.0694     & -0.1358      & 0.3495    & -2.0068   \\ \hline
101   & 12220                  & 0.1168 & -0.0291    & -0.2704      & 0.2942    & -1.9165   \\ \hline
\end{tabular}
\caption{Comparison of $R^{2}$ for gas used estimators}
\label{table:gas-usage-r2}
\end{table}

\begin{figure*}[ht]
    \centering
    \includegraphics[width=370px]{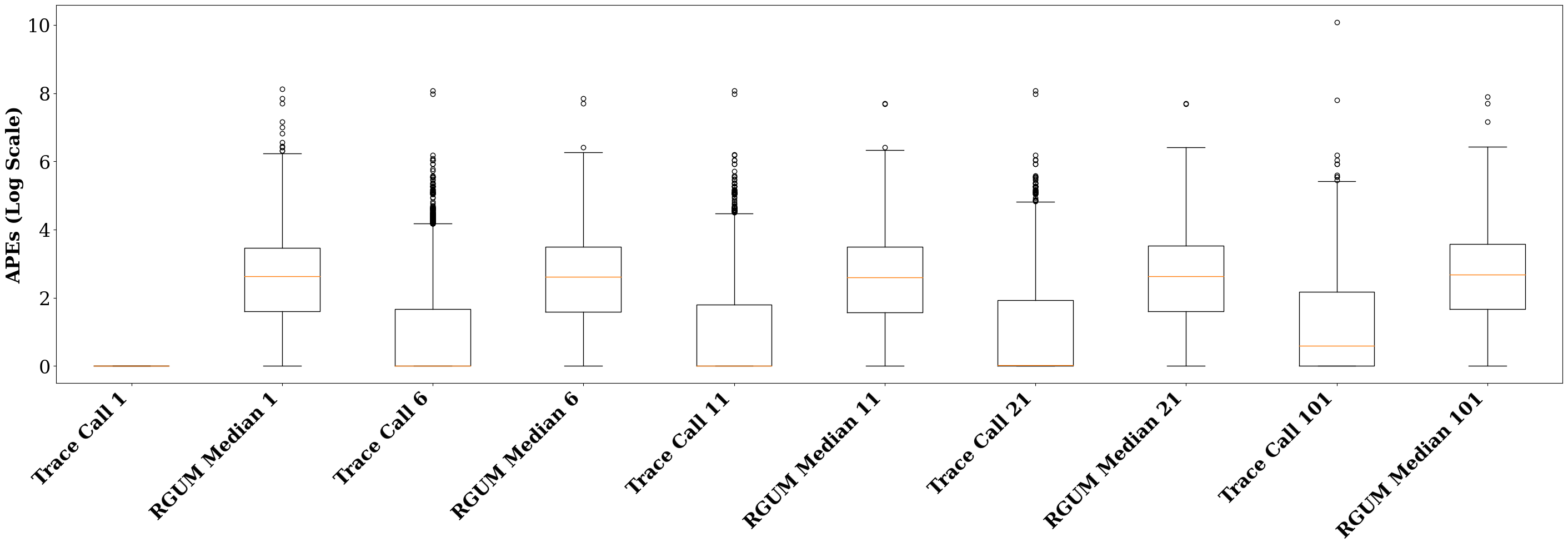}
    \caption{Boxplot of \glspl{ape} for gas used Estimates in Transactions with Block-State-Dependent Opcodes}
    \label{image:boxplot-gas-usage-bad-opcodes}
\end{figure*}

Table \ref{table:gas-usage-r2} compares the precision of the estimators according to the $R^{2}$ metric. In it, the results obtained by \textproc{TraceCall} ranged from 1 to 0.9966 in dataset $D_{1}$ and from 1 to 0.1168 in dataset $D_{2}$. In dataset $D_{1}$, among the versions of \gls{rgum}, \gls{rgum-min} presented the most balanced $R^{2}$ values, increasing from 0.9076 when $\Delta = 1$ to 0.9268 when $\Delta = 101$. Once again, \gls{rgum-max} obtained the best results in dataset $D_{2}$, with $R^{2}$ values ranging from 0.2817 to 0.3557. This table also reports the only case where \textproc{TraceCall} performed worse than \gls{rgum-max}. Specifically, when $\Delta = 101$ in dataset $D_{2}$, the $R^{2}$ of \textproc{TraceCall}'s estimates was 0.1168, while \gls{rgum-max} reached an $R^{2}$ of 0.2942. We investigated the reason for this unusual result and found that, as with three versions of \gls{rgum} in dataset $D_{1}$ with $\Delta = 11$, the $R^{2}$ of \textproc{TraceCall} was also significantly impacted by a single transaction, as illustrated later in Figure \ref{image:boxplot-gas-usage-bad-opcodes}.

\begin{figure*}[ht]
    \centering
    \includegraphics[width=370px]{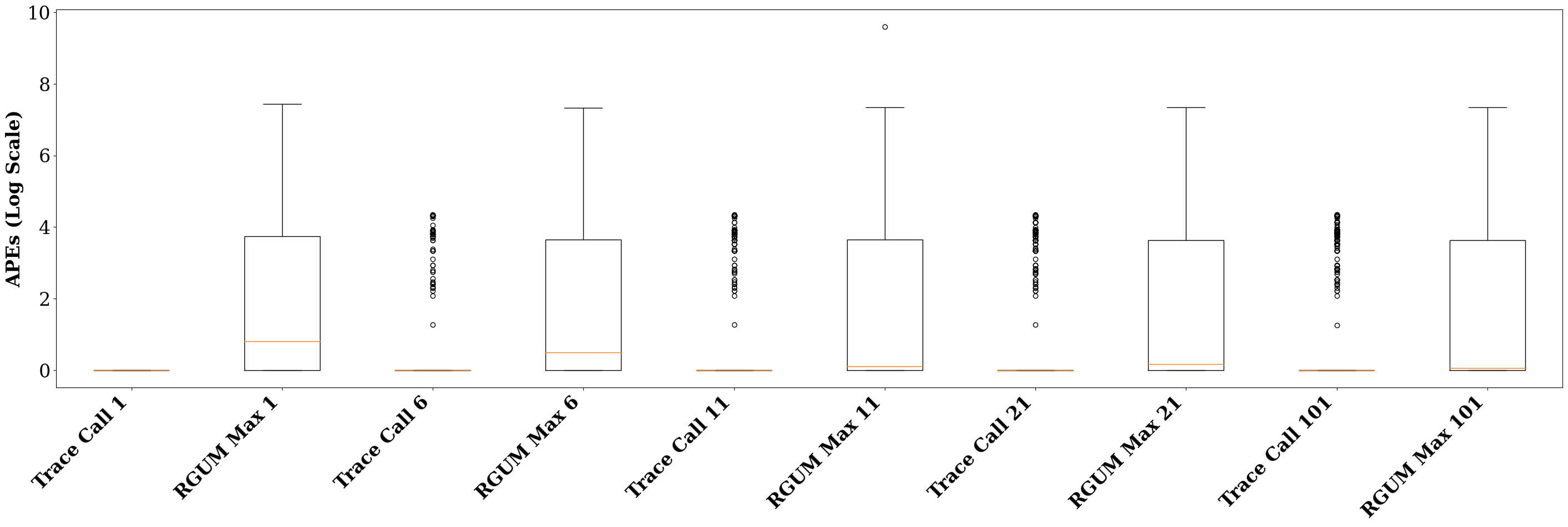}
    \caption{Boxplot of \glspl{ape} for gas used Estimates in Transactions without Block-State-Dependent Opcodes}
    \label{image:boxplot-gas-usage-good-opcodes}
\end{figure*}

Figure \ref{image:boxplot-gas-usage-good-opcodes} presents the distributions of \glspl{ape} obtained by \textproc{TraceCall} and \gls{rgum-max} in estimating gas used by transactions in Dataset $D_{1}$. Figure \ref{image:boxplot-gas-usage-bad-opcodes} shows the distributions of \glspl{ape} from \textproc{TraceCall} and \gls{rgum-median} in Dataset $D_{2}$. In both datasets, \textproc{TraceCall} performed similarly to \textproc{EstimateGas} in Experiment $E_{1}$, except when $\Delta = 101$ in Dataset $D_{2}$. For this particular case, \textproc{TraceCall} presented an unusually high \gls{ape} value for a single transaction.
As mentioned earlier, this outlier significantly impacted \textproc{TraceCall}'s $R^{2}$ in this scenario.
Looking at the \textproc{TraceCall} boxplots in both datasets, we can draw conclusions similar to those obtained regarding the precision of \textproc{EstimateGas}:\textproc{TraceCall} accurately estimated gas usage for 50\% of the transactions in Dataset $D_{1}$, regardless of $\Delta$ values. Furthermore, except for $\Delta = 21$ and $\Delta = 101$, \textproc{TraceCall} estimated the exact gas usage for 25\% of the transactions in Dataset $D_{2}$. When comparing with \gls{rgum-max} in Dataset $D_{1}$ and \gls{rgum-median} in Dataset $D_{2}$, all \textproc{TraceCall} outliers were below the upper whiskers of the boxplots of the former, and the first and third quartiles of all its boxplots remained below the median of the latter, reinforcing the superiority of \textproc{TraceCall} in both cases.

\begin{figure*}[ht]
    \centering
    \includegraphics[width=370px]{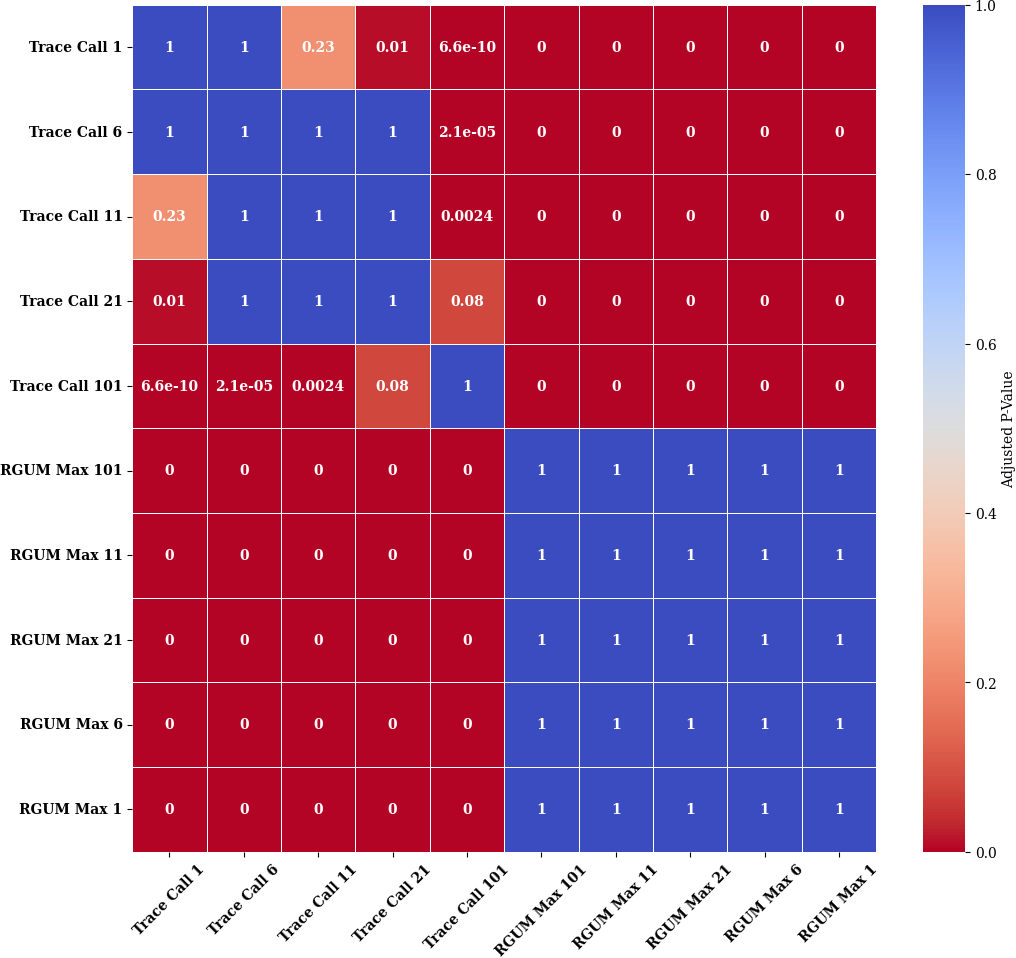}
    \caption{Heatmap of \glspl{ape} for gas used Estimates in Transactions without Block-State-Dependent Opcodes}
    \label{image:heatmap-gas-usage-good-opcodes}
\end{figure*}

\begin{figure*}[ht]
    \centering
    \includegraphics[width=370px]{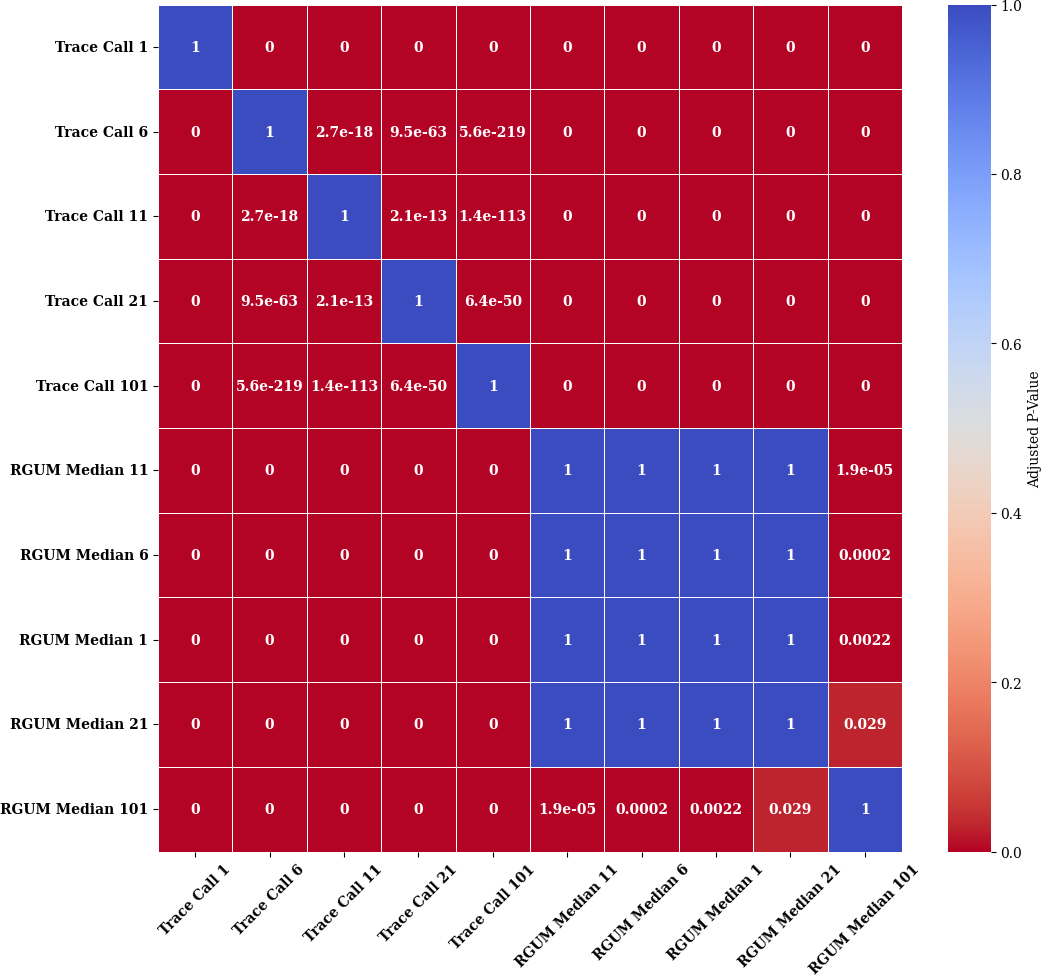}
    \caption{Heatmap of \glspl{ape} for gas used Estimates in Transactions with Block-State-Dependent Opcodes}
    \label{image:heatmap-gas-usage-bad-opcodes}
\end{figure*}

Figures \ref{image:heatmap-gas-usage-good-opcodes} and \ref{image:heatmap-gas-usage-bad-opcodes} present the heatmaps resulting from the application of the \textproc{Kruskal-Wallis} test, along with the \textproc{Conover} post hoc test, to statistically compare the precision of \textproc{TraceCall} against \gls{rgum-max} in Dataset $D_{1}$, and \textproc{TraceCall} against \gls{rgum-median} in Dataset $D_{2}$. Both figures demonstrate results similar to those of Experiment $D_{1}$. In both datasets, the worst-case precision of \textproc{TraceCall}, specifically when $\Delta = 101$, was statistically superior to the best case of \gls{rgum-max} in Dataset $D_{1}$ and \gls{rgum-median} in Dataset $D_{2}$. Considering a significance level of 5\%, there was no statistical difference between the results obtained by \textproc{TraceCall} with $\Delta = 1$, $\Delta = 6$, and $\Delta = 11$ in Dataset $D_{1}$. Here, unlike in Experiment $E_{1}$, even when considering a significance level of 1\%, the statistical test reports a significant difference between the precision of \textproc{TraceCall} when $\Delta = 1$ and $\Delta = 21$. In Dataset $D_{2}$, which contains transactions with block-state-dependent opcodes, the precision of \textproc{TraceCall} significantly decreased as the $\Delta$ value increased, similarly to what happened with \textproc{EstimateGas} in Experiment $E_{1}$, reinforcing our understanding that such opcodes also affect the gas used by a transaction.
Based on these results, we reject the null hypothesis for Experiment $E_{2}$ ($H_{0, E_{2}}$).

\subsection{Minimum Gas Limit vs Gas Used Experiment ($E_3$)} \label{section:minimum-gas-limit-vs-gas-used-experiment}

Following the protocol defined in Section \ref{subsection:experimental-design}, we present the results of Experiment $E_{3}$ in this subsection. Below, we describe the outcomes of the comparison between the distribution of the minimum gas limit of the analysed transactions and the distribution of its respective gas used.

Tables \ref{table:metrics-good-opcodes} and \ref{table:metrics-bad-opcodes} list, respectively, the median \glspl{ape}, mean \glspl{ape}, standard deviations of the \glspl{ape}, and $R^{2}$ of the gas cost distributions for the transactions in  Dataset $D_{1}$ and Dataset $D_{2}$. In Dataset $D_{1}$, the median \gls{ape} was equal to 0, indicating that in at least 50\% of the transactions, there is no difference between the minimum gas limit and the gas used. Conversely, in Dataset $D_{2}$, the median \gls{ape} was 6.08, suggesting that even for transactions with block-state-dependent opcodes, in at least 50\% of the cases, the difference between the minimum gas limit and the gas used is small. However, when comparing the mean \glspl{ape} shown in the two tables, it is evident that the difference between the distributions of the minimum gas limit and the gas used in Dataset $D_{2}$ is considerably greater than that in Dataset $D_{1}$. The $R^{2}$ values of the Table \ref{table:metrics-good-opcodes} reveal a linear relationship between the distributions of minimum gas limit and gas used in the transactions of Dataset $D_{1}$, as the $R^{2}$ for this dataset was equal to 1. However, the $R^{2}$ associated with Dataset $D_{2}$ was negative ($R^{2} = -3.79$), indicating the impossibility of determining the values of minimum gas limit based on the gas used by the transactions in this dataset.

\begin{figure*}[ht]
    \centering
    \includegraphics[width=370px]{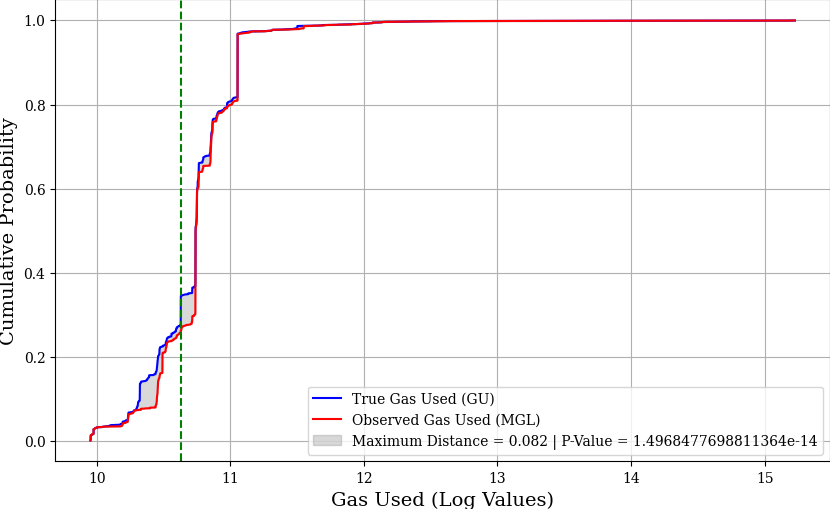}
    \caption{\gls{ks} Plot of minimum gas limit and gas used by the Transactions without Block-State-Dependent Opcodes}
    \label{image:ks-good-opcodes}
\end{figure*}

\begin{figure*}[ht]
    \centering
    \includegraphics[width=370px]{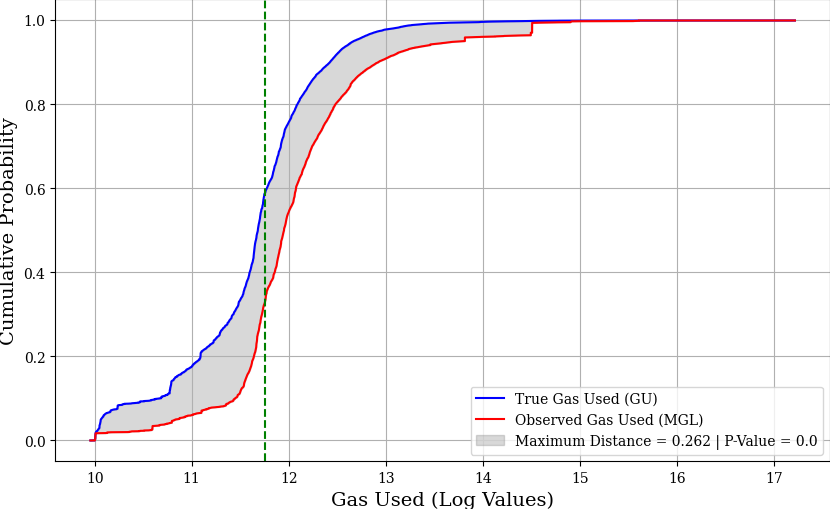}
    \caption{\gls{ks} Plot of minimum gas limit and gas used by the Transactions with Block-State-Dependent Opcodes}
    \label{image:ks-bad-opcodes}
\end{figure*}

Figures \ref{image:ks-good-opcodes} and \ref{image:ks-bad-opcodes} present the \gls{ks} plot for the comparison between the minimum gas limit and gas used distributions of the transactions belonging to Dataset $D_{1}$ and Dataset $D_{2}$, respectively. In both cases, we can see that the curves with the distributions of the accumulated frequencies of the minimum gas limit of the transactions assume a shape with values consistently greater than those of gas used. This difference is evidenced by the values of the maximum distances between the two curves in each of the plots.

In Figure \ref{image:ks-good-opcodes}, the maximum distance between the two distributions is 0.082. In Figure \ref{image:ks-bad-opcodes}, this value is 0.262. In both figures, we observe that the \textproc{p-value} resulting from the statistical comparison between the distributions of minimum gas limit and gas used by transactions was nearly negligible, being $1.5 \times 10^{-14}$ in the first and approximately 0 in the second. These results indicate that, under any commonly used significance level, there is a significant difference between the distributions of minimum gas limit and gas used in both analysed datasets. Based on these results, we reject the null hypothesis for Experiment $E_{3}$ ($H_{0, E_{3}}$).

\begin{table}[ht]
\centering
\resizebox{200px}{!}{%
\begin{tabular}{|c|c|c|c|c|}
\hline
\multirow{2}{*}{\textbf{Gas Used Metrics}} & \textbf{Median} & \textbf{Mean} & \textbf{SD} & \textbf{R$^{2}$} \\
                                  & 0             & 2.91        & 5.69     & 1      \\ \hline
\end{tabular}%
}
\caption{Minimum gas limit and gas used metrics of the Transactions without Block-State-Dependent Opcodes}
\label{table:metrics-good-opcodes}
\end{table}

\begin{table}[ht]
\centering
\resizebox{200px}{!}{%
\begin{tabular}{|c|c|c|c|c|}
\hline
\multirow{2}{*}{\textbf{Gas Used Metrics}} & \textbf{Median} & \textbf{Mean} & \textbf{SD} & \textbf{R$^{2}$} \\
                                  & 6.08            & 297.27        & 1406.36     & -3.79       \\ \hline
\end{tabular}%
}
\caption{Minimum gas limit and gas used metrics of the Transactions with Block-State-Dependent Opcodes}
\label{table:metrics-bad-opcodes}
\end{table}

\subsection{Discussion} \label{subsection:discussion}

In this subsection, we analyse the results obtained from experiments $E_{1}$, $E_{2}$, and $E_{3}$, with the aim of answering the research questions formulated in Section \ref{section:introduction}.

\textbf{\gls{rq1}}: \textbf{Under what circumstances is the minimum gas limit of a transaction at block $t$ expected to be a near-perfect estimation at block $t + \Delta$?}

To answer this research question, it is important to analyse how the precision of the \textproc{EstimateGas} function decreased as the $\Delta$ value increased in the two analysed datasets. Furthermore, it is crucial to interpret to what extent the presence of block-state-dependent opcodes in transactions contributes to this precision loss.

The results of Experiment $E_{1}$ revealed that in both analyzed datasets, the medians of the \glspl{ape} obtained by \textproc{EstimateGas} were 0, regardless of the $\Delta$ value, with the exceptions of $\Delta = 21$ and $\Delta = 101$ in Dataset $D_{2}$, where the medians of their \glspl{ape} were 0.01 and 0.55, respectively. These results are particularly interesting because they show that even when $\Delta = 101$, \textproc{EstimateGas} managed to maintain accuracies of 100\% and 99.45\% in at least half of the transactions in Dataset $D_{1}$ and Dataset $D_{2}$, respectively. The boxplots of the distributions of the \textproc{EstimateGas} \glspl{ape} further demonstrated that the percentage of transactions where this function maintained 100\% accuracy reached at least 75\% in Dataset $D_{1}$.

When considering both the mean \glspl{ape} and the $R^{2}$ of their estimates, we see that the precision of \textproc{EstimateGas} decreased as the $\Delta$ value increased. In Dataset $D_{1}$, this decline was considerably smaller, and the application of the statistical test did not indicate significant differences between the precision of \textproc{EstimateGas} when $\Delta = 1$, $\Delta = 6$, and $\Delta = 11$. In Dataset $D_{2}$, the precision loss was more pronounced, and the application of the statistical test revealed significant differences in the precision of \textproc{EstimateGas}, even when the $\Delta$ value increased from 1 to 6.

At the time of writing, the time to insert a new block into the Ethereum blockchain is approximately 12 seconds \cite{Ethereumdoc}. Given this average time, the results of Experiment $E_{1}$ suggest that \textit{eth\_estimateGas()} maintains precision similar to that of a perfect estimator when used to estimate the minimum gas limit of transactions yet to be inserted into the Ethereum blockchain, provided that the following conditions are met: the transaction must not contain block-state-dependent opcodes and must be inserted into the blockchain within 2 minutes of its minimum gas limit being estimated. If any of these conditions is not met, it is expected that, as the $\Delta$ value increases, there will be a gradual deviation from what would be considered a precision similar to that of a perfect estimator, with significantly greater precision losses when dealing with transactions containing block-state-dependent opcodes.

\textbf{\gls{rq2}}: \textbf{Under what circumstances is the gas used by a transaction at time $t$ expected to be a near-perfect estimation at block $t + \Delta$?}

Similar to what happened with \gls{rq1}, the answer to this research question depends on analysing the precision of the \textproc{TraceCall} function at different $\Delta$ values and interpreting how much the precision of this function was affected by the presence of block-state-dependent opcodes.

The analysis of the results obtained by \textproc{TraceCall} in Experiment $E_{2}$ revealed a precision similar to that of \textproc{EstimateGas} in Experiment $E_{1}$. Considering their median \glspl{ape}, we see that the only instances in which \textproc{TraceCall} did not achieve a median \gls{ape} of 0 were when $\Delta = 21$ and $\Delta = 101$ in Dataset $D_{2}$, where the results were 0.02 and 0.8, respectively. This indicates that \textproc{TraceCall} estimated the exact gas used value in at least 50\% of the transactions in Dataset $D_{1}$, while maintaining an accuracy of 99.2\% in at least half of the transactions in Dataset $D_{2}$. The boxplots of \textproc{TraceCall}'s \gls{ape} distributions demonstrated that in Dataset $D_{1}$, the percentage of transactions where \textproc{TraceCall} estimated the exact gas used value was at least 75\%.

As the Delta value increased, both the mean \glspl{ape} and $R^{2}$ of the \textproc{TraceCall} estimates consistently worsened across the two datasets analysed. Similar to what occurred with \textproc{EstimateGas} in Experiment $E_{1}$, the application of the statistical test indicated no significant differences in the \textproc{TraceCall} precision when $\Delta = 1$, $\Delta = 6$, and $\Delta = 11$ in Dataset $D_{1}$. In Dataset $D_{2}$, the \textproc{TraceCall} precision decreased significantly even when the value of $\Delta$ increased from 1 to 6.

Based on the results of Experiment $E_{2}$, we argue that \textit{debug\_traceCall()} maintains precision similar to that of a perfect estimator when used to estimate the gas used by transactions yet to be inserted into the Ethereum blockchain, provided that the same conditions required by \textit{eth\_estimateGas()} when estimating the minimum gas limit are met: the transaction must not contain block-state-dependent opcodes and must be inserted into the blockchain within 2 minutes of its gas used being estimated. A gradual deviation from what would be considered similar precision to a perfect estimator is expected if any of these conditions are not met, particularly for transactions containing block-state-dependent opcodes, where \textit{debug\_traceCall()} suffered more significant precision losses.

\textbf{\gls{rq3}}: \textbf{Are there significant differences between the gas used and minimum gas limit of Ethereum \gls{sc} call transactions?}

Considering the minimum gas limit of the analysed transactions as observed gas used and the gas used as their respective true gas used, Experiment $E_{3}$ demonstrated that the median \gls{ape} of the gas used distribution for transactions in Dataset $D_{1}$ was 0. This indicates that, in at least 50\% of the transactions in this dataset, there is no difference between its minimum gas limit and its respective gas used. In Dataset $D_{2}$, the median \gls{ape} was 6.08, revealing that, even in a set of transactions with block-state-dependent opcodes, the difference between the minimum gas limit and the gas used was relatively small in at least half of the cases. However, the discrepancy between the mean \gls{ape} values of the two datasets suggests that the difference between the minimum gas limit and the gas used by the transactions in Dataset $D_{2}$ is considerably larger than in Dataset $D_{1}$. Furthermore, the $R^{2}$ value obtained in Dataset $D_{1}$ ($R^{2} = 1$) revealed a linear relationship between the minimum gas limit and the gas used by the transactions in this dataset, suggesting that, in a set of transactions without block-state-dependent opcodes, it is possible to predict the minimum gas limit of a transaction by knowing its gas used, and vice versa. On the other hand, the negative $R^{2}$ value for Dataset $D_{2}$ ($R^{2} = -3.79$) indicates the impossibility of predicting minimum gas limit based on the gas used by transactions that contain block-state-dependent opcodes.

The statistical comparison rejected the hypothesis of equality between the distributions of minimum gas limit and gas used by transactions in the two analysed datasets. This indicates that, although the median \glspl{ape}, mean \glspl{ape}, and $R^{2}$ values obtained in this experiment suggest that the differences between the observed gas used and the true gas used of transactions with block-state-dependent opcodes tend to be substantially larger than in cases where such opcodes are absent, significant differences between the values of minimum gas limit and gas used are expected in both types of transactions.

\subsection{Threats to Validity} \label{section:threats-to-validity}

In this subsection, we discuss the main threats to the validity of our experimental results. We identify these threats using the guidelines defined by Wohlin et al. \cite{Wohlin:2012}.

\subsubsection{Internal Validity Threats} \label{subsubsection:internal-validity-threats}

At the time of writing, the Ethereum blockchain contains more than 2 million blocks, each with multiple transactions. Considering the infeasibility of analyzing all transactions contained in the blockchain, we selected only transactions belonging to the Ethereum Mainnet's \textit{Bellatrix} fork for our experiment. This selection, although made to achieve the best balance between the number of blocks available in our sample and the computational cost of processing and locally storing the transactions in a format compatible with our analysis, constitutes a threat to the internal validity of our experiments, as it may introduce an unwanted bias into the results.

In our experiments, we assume that the state of the \gls{evm} when processing a zero-index transaction in a given block $B$ is equal to the final state of block $B - 1$. Especially in cases of transactions that depend on specific block elements, such as the hash, block number, or timestamp, this assumption, although with minimal impacts, may introduce an uncontrolled source of variation, and therefore represent an additional threat to internal validity.

A third threat to internal validity arises from the lack of availability of the source code for most of the gas estimators proposed in the current literature. Given this limitation, we reduced the scope of experiments $E_{1}$ and $E_{2}$ to the comparison of the \textproc{EstimateGas} and  \textproc{TraceCall} functions against four versions of \gls{rgum}. Because of this, the results obtained in experiments $E_{1}$ and $E_{2}$ may be biased, as they do not consider the other gas estimators currently proposed.

\subsubsection{External Validity Threat} \label{subsubsection:external-validity-threat}

To the best of our knowledge, no fork subsequent to \textit{Bellatrix} has introduced any significant changes to the gas system that would invalidate the results obtained in this work. However, considering the dynamic nature of Ethereum, the choice of this single fork represents a threat to the external validity of our experiments. This is because new protocol updates that result in consistent changes to the gas system may render the generalization of the results obtained here unfeasible.

\subsubsection{Construct Validity Threat} \label{subsubsection:construct-validity-threat}

Executing the zero-index transaction of block $B$ under the context of the final state of block $B - 1$ may generate subtle discrepancies compared to executing the same transaction in its actual context, particularly in cases where this transaction is strongly influenced by specific attributes of $B$, such as its hash, number, and timestamp. Thus, the assumption that the final state of block $B - 1$ is equal to the initial state of block $B$ represents a threat to the construction validity of our experiments, as, even though it occurs only in very specific contexts, such an assumption may generate results that distort reality.

\subsubsection{Conclusion Validity Threat} \label{subsubsection:conclusion-validity-threat}

The comparison between the precision of \textproc{EstimateGas} and  \textproc{TraceCall} functions and four versions of \gls{rgum}, due to the unavailability of the source code of other gas estimators identified during the preparation of this article, represents a threat to the conclusion validity of the experiments $E_{1}$ and $E_{2}$, as the inclusion of the estimators that were necessarily disregarded in these experiments could significantly affect the results, especially if any of them exhibited significantly higher or lower accuracy than the estimators compared here.

\section{Related Work} \label{section:related-work}

In this section, we consider some of the main works available in the literature that are related to the Ethereum gas system. We split these works into two categories: gas estimation and gas analysis.

\subsection{Gas Estimation} \label{gas-estimation}

The \gls{solc} \cite{solc} is the official compiler for the Solidity programming language \cite{Solidity} and is primarily designed to convert source code written in Solidity into \gls{evm} bytecode. This compiler has an additional functionality that allows it to estimate the gas used by functions contained in the \glspl{sc} it compiles.
Operating as an offline gas estimator, \gls{solc} performs its estimates without considering transaction arguments and blockchain states. Consequently, this compiler cannot effectively handle functions whose gas used is influenced by these factors. This limitation is especially evident when estimating the gas used by functions that contain dynamic loops. In such cases, \gls{solc} is limited to always returning \textit{infinity}.

Marescotti et al. \cite{Marescotti:2018} proposed two estimators that predict the gas used by \gls{sc} functions via the identification of their most expensive execution paths. To do so, they created a \gls{gcg} model based on traditional \glspl{cfg}. These \glspl{gcg} have additional edges and nodes to represent the function arguments and the state variables. Both methodologies identify potentially divergent \glspl{gcg} using bounded model checking techniques with \gls{smt} solvers, which rely on manually adjusting the loop unwinding threshold to return accurate results.
 
\gls{gastap} \cite{Albert:2018}, later published as \gls{gasol} \cite{Albert:2020}, yields the worst-case gas used by all public functions of each \gls{sc} analysed.
To perform its estimations, \gls{gastap} first generates \glspl{cfg}, followed by creating rule-based representations. These representations are then converted into size relations, which are used to formulate the gas equations. Based on these equations, this estimator finally returns the worst-case gas used by the functions.

Although \gls{gastap}, \gls{gasol}, and the two methods proposed by Marescotti et al. demonstrated greater accuracy than \gls{solc} in some specific scenarios, all four of these estimators, being offline, disregard transaction arguments and blockchain states. Like \gls{solc}, these estimators also have a severely impacted accuracy when dealing with functions whose gas used values vary according to these factors.

\gls{tdge} \cite{Li:2020} applies machine learning algorithms to estimate the gas used by transactions based on their gas used history. This estimator is specifically designed to handle transactions executed by functions that contain loops and have their gas costs dynamically changed according to the values of their arguments and the state of the blockchain. To estimate the gas used by a transaction sent to a function $F$ of a given \gls{sc}, \gls{tdge} first discovers the block $\beta$ where the \gls{sc} was created. It then creates a local fork of the Ethereum Mainnet from $\beta - 1$ and trains its machine learning algorithms tracking the gas used by executing multiple transactions directed to $F$, with different variations in the argument values.

Trained from a local fork of the Ethereum Mainnet, starting at block $\beta - 1$, this estimator does not have access to the transactions that interacted with the function $F$ on the Ethereum Mainnet. As a result, although its training process involves executing various transactions directed at $F$, with different argument values, there is no guarantee that these transactions will faithfully simulate the different contexts to which $F$ was exposed on the real blockchain. Consequently, the estimator may not effectively generalize the knowledge obtained during training when handling real-world transactions.

\textproc{V-GAS} \cite{Ma:2022} was proposed as an online gas used estimator based on static analysis and feedback-driven fuzz testing. Taking as input a binary file (.bin) along with the application binary interface (.abi) of a given \gls{sc}, this estimator aims to predict the gas used by all its public functions.
To estimate the gas used by an \gls{sc}, \textproc{V-GAS} uses a version of the \gls{evm} written in JavaScript (JS-EVM) to create multiple instances of the contract on a local blockchain, assigning random values to all its state variables. Then, for each function of the instantiated contracts, it creates a series of transactions with randomly assigned argument values. Finally, using a genetic algorithm-based optimization technique, \textproc{V-GAS} identifies the values of state variables and arguments that result in the worst-case gas used by each function.

By identifying only the worst-case gas used by the analysed functions, \textproc{V-GAS} operates similarly to offline estimators but adds the ability to effectively handle functions whose gas used is highly dependent on transaction arguments and blockchain states. However, by suggesting only a single value for the gas used by each function, this approach, like all offline estimators, will in most cases lead to an overestimation of gas, which, as mentioned in Section \ref{subsection:estimation-of-minimum-gas-limit}, can result in a few practical problems.

Zarir et al. \cite{Zarir:2021} conducted an exhaustive investigation into the gas used by Ethereum transactions during the Byzantine era and, considering the set of \gls{sc} functions that received at least 10 transactions, identified that only 25\% of them exhibited instability in the gas used. Based on this, they inferred that, in most cases, the history of the gas used by the functions follows stable and reliable patterns.

\gls{rgum} is a gas used estimator derived from the research of Zarir et al \cite{Zarir:2021}. Designed assuming that most transactions have a stable gas used history, this estimator predicts the gas used by transactions based on their histories. More specifically, \gls{rgum} assumes that the gas used by a given transaction that invokes a function $F$ is equal to the mean gas used by the last 10 transactions sent to $F$. We have used \gls{rgum} with several variations in our experiments and have analysed their precision.

When predicting the gas used by \gls{sc} functions based solely on the mean of the last 10 transactions directed at these functions, \gls{rgum}, although effective in cases where the level of gas used by the functions does not vary significantly, proves to be highly ineffective for functions whose gas used is strongly influenced by transaction arguments and blockchain states, which are inherently dynamic.

Although our work falls into the category of studies on gas estimation, it differs from the others in some aspects. We propose minimum gas limit as a metric distinct from gas used; we present a detailed exposition with examples where we introduce scenarios in which the values of these metrics may diverge. 
Moreover, as part of our experimental study, we have demonstrated that these two metrics typically diverge in practice. Furthermore, we have demonstrated that contract executions can be qualitatively clustered by some opcodes associated with how dynamic the function executions can be. Finally, we have reported how, for a given transaction, calculations for the minimum gas limit and the gas used can serve as near-perfect estimations for these values if the transaction is included in a block not so far after the one where the calculation was made.

\subsection{Gas Analysis} \label{subsection:gas-analysis}

The current literature contains several proposed works to analyse the Ethereum gas system with different objectives. Perez-Carrasco et al. \cite{Perez-Carrasco:2020} performed a mapping of the gas used by Ethereum transactions to the usage of hardware resources. In their work, Bistarelli et al. \cite{Bistarelli:2019} and Khan et al. \cite{Khan:2021} analysed, respectively, the execution frequency of single opcodes and the correlation between opcode execution and source code features. Regarding gas used profiling, Ding et al. \cite{Ding:2020}, Albert et al. \cite{Albert:2020}, and Signer et al. \cite{Signer:2018} presented, respectively, results based on single opcodes, Ethereum yellow paper categories, and source code line level. In addition, Ashraf et al. \cite{Ashraf:2020} performed gas used profiling based on fuzz testing, Correas et al. \cite{Correas:2021} established theoretical upper-bounds for resource consumption, and Severin et al. \cite{Severin:2022} proposed higher-level gas cost categories to serve as a bridge between the technical opcodes and the design pattern literature.

Some works also analyse the relationship between the gas used by Ethereum transactions and security vulnerabilities, and how to optimise of the source code of \glspl{sc} to reduce the gas cost of their functions. Luu et al. \cite{Luu:2016} proposed \textit{OYENTE}, a framework designed to discover security bugs in \glspl{sc} based on symbolic execution. In his thesis, Kong \cite{Kong:2017} introduced several types of gas vulnerabilities and presented algorithms that compute loop bounds to identify them. Based on Kong's work, Grech et al. \cite{Grech:2018} identified and classified gas-focused vulnerabilities and proposed a static program analysis technique named \textit{MadMax} to automatically detect them. Chen et al. \cite{Chen:2017} identified seven code patterns in \glspl{sc} associated with high values of gas used and developed a symbolic execution-based tool capable of discovering gas-costly patterns in bytecode. In their research, Marchesi et al. \cite{Marchesi:2020} documented 24 design patterns that aim to help save gas in the development of \glspl{sc} on Ethereum.

By proposing the minimum gas limit and gas used as two distinct concepts and identifying a list of opcodes that, being dependent on block states, significantly influence the estimation of these two metrics, this work, although primarily focused on gas estimation issues, also brings new perspectives to the current body of research directed towards gas analysis. In this work, for example, we demonstrated that out-of-gas exceptions, related to security issues \cite{Grech:2018} such as denial-of-service attacks, freezing of funds in smart contracts, and gas griefing, are also directly associated with the concept of minimum gas limit, and not with gas used. By understanding that these metrics can have distinct values, new research exploring the relationship between gas and security in smart contracts will be able to establish more precise associations between these two topics. Furthermore, new studies exploring the relationship between opcodes and gas can benefit from the results presented in this paper by considering the influence that block-state-dependent opcodes exert on the predictability of the gas used by transactions.

\section{Conclusions}\label{section:conclusions}

The main contributions of this paper are related to a detailed and systematic analysis of the Ethereum gas system. Particularly, we propose a precise notion of the \emph{minimum gas limit} of a transaction, a qualitative and quantitative study of the differences between this concept and the gas used by a transaction, and present an online estimator for each of these metrics.

As far as we are aware, this paper is the first to propose a precise notion of minimum gas limit. In fact, other papers in this area are unaware of the distinction between this concept and gas used, and conflate both~\cite{Liu:2020,Zarir:2021}; these works assert that by predicting the gas used by a transaction, this value can be used to set its gas limit. In contrast, we provide evidence that, for many cases, this strategy does not work, and present a detailed analysis of the concepts of minimum gas limit and gas used, focusing on the cases where these two concepts diverge. Moreover, after this qualitative exposition, we present an empirical study analysing a fragment of Ethereum's history that shows that there are statistically significant differences between these two metrics; in practice, these two values often diverge for a transaction. Thus, our work serves as both a warning and as documentation to Ethereum application developers and users on the (existence of the concept of) minimum gas limit and gas used, and how they differ. 

We also propose an estimator for each of these metrics. Given the conflation of these concepts, the estimators in the literature predict only gas used. We analyse whether, for a given transaction, the minimum gas limit at the Ethereum blockchain state (after block) $t$ can reliably estimate the minimum gas limit at block $t + \Delta$. Our experimental study shows that this estimator is practically perfect for $\Delta \leq 11$ for less dynamic transactions, and it is still fairly precise for large values of $\Delta$ and more dynamic types of transactions; the dynamicity of a transaction relates to the block-state-dependent opcodes. We carry out the same analysis for gas used reaching the same result. Hence, we have provided two strategies that can reliably estimate the gas used and the minimum gas limit. These estimators can be conveniently implemented using the Ethereum client functions \textit{eth\_estimateGas()} and \textit{debug\_traceCall()}. These estimators can help tackle many of the issues associated with the gas system of Ethereum, like the unpredictability of gas costs, which is one of the main obstacles to the widespread adoption of \glspl{dapp}\cite{AlShamsi:2022}. They can also help with the optimisation of gas usage by contracts and with the analysis of security risks associated with excessive gas consumption.

We also emphasise some limitations of our work. As part of our experimental project, we presented a list of block-state-dependent opcodes; throughout the experiments, we reported results suggesting the influence of such opcodes on the accuracy of gas estimators. Although we identified this relationship between block-state-dependent opcodes and the precision of gas estimators, we did not conduct a fine-grained empirical study on how each of these opcodes can individually influence gas estimates.

Furthermore, we selected transactions belonging to a single fork that were allocated at the zero index position in their respective blocks. This measure ensured that all selected transactions were included in the blockchain according to the same consensus rules and allowed us to associate the state of the \gls{evm} in a block of height $b - 1$ with a transaction selected from a block $b$. Considering that the Ethereum blockchain currently accumulates more than 2 million blocks, each with multiple transactions, one should be aware that the scope of the experiments conducted in this work, although containing tens of thousands of transactions, represents a relatively small percentage of the total transactions stored on the Ethereum blockchain. We have, however, no reason to believe that our empirical conclusions would not apply to the entire history of Ethereum; as far as the gas system is concerned, there is no significant difference between the fragment of the Ethereum history that we analyse and the overall history.

The results and limitations presented in this work can drive various future research directions. We present some of them below:

\begin{itemize}

    \item Detailed empirical analysis of the behaviour of block-state-dependent opcodes, with the aim of understanding how each of these opcodes can individually impact the precision of gas estimators.

    \item Expansion of the experimental scope, including transactions from other forks, with varied indices and positions within blocks, while also considering different levels of delay in the inclusion of these transactions on the blockchain.

    \item Development of adaptive mechanisms, based on machine learning or other approaches, aiming to both improve the precision of gas estimators in transactions containing block-state-dependent opcodes and maintain this precision even in the presence of delays in the inclusion of these transactions on the blockchain.

    \item In-depth investigation into the relationship between the minimum gas limit and gas used by transactions, especially in cases where these transactions do not contain block-state-dependent opcodes, with the goal of proposing dual-approach gas estimators capable of deriving the minimum gas limit estimate from the gas used by transactions and vice versa.
\end{itemize}

\backmatter

\bmhead{Acknowledgements}
We would like to thank Waldemar Ferreira Neto for fruitful discussions concerning the experimental design. This project is partially supported by CNPq, grant 443588/2023-6.

\bibliography{sn-bibliography}

\end{document}